\documentstyle[aps,prb,multicol,epsfig]{revtex}
\title{Onsager Loop-Transition and First Order Flux-Line Lattice
Melting in High-$T_c$ Superconductors}
\author{A. K. Nguyen and A. Sudb{\o} }
\address{Department of Physics\\
	 Norwegian University of Science and Technology,
	 N-7034 Trondheim, Norway \\}

\newcommand{\kp    }{\! +      \!}
\newcommand{\km    }{\! -      \!}
\newcommand{\keq   }{\! =      \!}
\newcommand{\kgt   }{\! >      \!}
\newcommand{\klt   }{\! <      \!}
\newcommand{\ksim  }{\! \sim   \!}
\newcommand{\ksimeq}{\! \simeq \!}
\newcommand{\bleq}{ \end{multicols}
		    \vspace*{-3.5ex}
		   {\tiny
                    	\noindent
	            	\begin{tabular}[t]{c|}
                    		\parbox{0.493\hsize}{~} \\
		    		\hline
		    	\end{tabular}
		   }
                  }
\newcommand{\eleq}{{\tiny
			\hspace*{\fill}
			\begin{tabular}[t]{|c}
				\hline
                    		\parbox{0.49
				\hsize}{~} \\
                    	\end{tabular}
		   }
		    \vspace*{-2.5ex}
		    \begin{multicols}{2}
                  }

\begin{document}

\maketitle

\begin{abstract}
Monte-Carlo simulations in conjunction with finite-size scaling analysis
are used to investigate the $(H,T)$-phase diagram in uniaxial anisotropic
high-$T_c$ superconductors, both in zero magnetic field $(B \keq 0)$ and
in intermediate magnetic fields ($0 \! \ll \! B \! \ll \! B_{c2}$) for
various mass-anisotropies. The model we consider is the uniformly frustrated
anisotropic Villain Model, which is dual to the Lattice London Model with
an infinite London penetration length. The quantities we consider are various
helicity moduli, the structure function, the specific heat, and the
distribution of closed non-field induced vortex loops as a function of the
loop-size. In zero magnetic field, and for all anisotropies considered,
we find one single second order phase transition, mediated by an Onsager
vortex-loop unbinding transition, or blowout. This is the
superconductor-normal metal transition. A comparison with numerical
simulations and a critical scaling analysis of the zero-field loop-transition
yields the same exponent of the loop-distribution function at the critical
point. In the intermediate magnetic field regime, we find two anomalies in
the specific heat. The first anomaly at a temperature $T_m$ is associated
with the melting transition of the flux-line lattice. The Lindemann-ratio
at the melting is given by $c_L \approx 0.24$. The second anomaly at a
temperature $T_z$ is one where  phase coherence in the BCS order parameter
across the sample along the field direction is destroyed. We argue that
$T_m=T_z$ in the thermodynamic and continuum limit.
Hence, there is no regime where the flux-line lattice melts into a
disentangled flux-line liquid. The loss of phase coherence parallel to the
magnetic field in the sample is argued to be due to the proliferation of
closed non-field induced vortex loops on the scale of the magnetic length in
the problem, resulting in flux-line cutting and recombination. In the flux-line
liquid phase, therefore, flux-lines appear no longer to be well defined
entities. Above the melting temperature, the system always exhibits an
{\it incoherent vortex-liquid phase} characterized by lack of phase
coherence in the BCS order parameter parallel to the magnetic field.
For increasing anisotropy, we resolve a delta-function peak in the
specific heat. A finite-size scaling analysis of the delta-function peak
specific heat anomaly at the melting transition is used to extract the
discontinuity of the entropy at the melting transition. This entropy
discontinuity is found to increase rapidly with mass-anisotropy, at least
for not too layered compounds.
\end{abstract}

\begin{multicols}{2}
\section{Introduction}
A number of recent experiments have reported results of a first order melting
transition of the Abrikosov flux-line lattice (FLL)
\cite{Safar:L93,Pastoriza:L94,Doyle:L95,Zeldov:N95,Nishizaki:B96,Welp:L96,Schilling:N96,Junod:P96}. The reported magnitudes of the latent heat have
all in general been in surprisingly
good agreement with a {\it prediction} of Hetzel {\it et al.} for
the discontinuity in the entropy at the melting transition based on
extensive Monte-Carlo simulations of the uniformly frustrated $3D$
$XY$-model \cite{Hetzel:L92}. \\
Schilling {\em et al.} have reported calorimetric measurements on an
untwinned $YBa_{2}Cu_3 O_7$ (YBCO) single crystal, in the intermediate
field regime $B\!\!\in\!\![1-7]$ Tesla, and find a FLL melting transition
with a virtually field independent entropy jump $\Delta S \ksim 0.45 k_B$
per vortex per layer \cite{Schilling:N96}.
In $Bi_{2}Sr_{2}Ca_1 Cu_{2}O_{8}$ (BSCCO) single crystal at very low magnetic
inductions, $B\!\!\in\!\![1 \!-\! 375]$ Gauss (G), it was  found that a FLL
melting transition occurs with an enormous entropy jump
$\Delta S(B\!\!=\!\!1G) \ksim 6 k_B$ per vortex per layer
\cite{Zeldov:N95}. Furthermore, it was found that
$\Delta S(B)$ decreases for increasing B, and vanishes at $B \ksim 375$ G
\cite{Zeldov:N95}. It appears that
$\Delta S(B)$ increases dramatically only when $B \to 0, T \to T_c$.
In fact, it might be argued on the basis of the data of Zeldov {\em et al.},
that $\Delta S(B)$ {\it diverges} in this limit. \\
Recently, Te{\v s}anovi{\'c} \cite{Tesanovic:B95} and Nguyen {\em et al}.
\cite{Nguyen:L96} have proposed an explanation for the inordinately
large entropy jump found in Ref. \cite{Zeldov:N95}.
The idea is that the FLL melting transition takes place at roughly the
same temperature as a ``blowout'' of non-field induced degrees of freedom
involving closed vortex loops, resulting in a flux-line liquid phase with
considerably larger entropy than what the field-induced vortices alone
can provide. \\
As a step towards understanding these experimental results,  we carry
out extensive Monte Carlo simulations, together with a finite-size scaling
analysis, on the uniformly frustrated anisotropic Villain Model, to be
defined below.  This model will be argued to be appropriate for describing
the physics in extreme type-II superconductors such as the high-$T_c$
superconductors.  Here, we present a short review of our results. \\
For the case $B \keq 0$, we find  that the Villain Model has {\it one}
single second order phase transition of the $3DXY$-type for all anisotropies
considered. {\it The phase transition in zero magnetic field is caused
exclusively by a vortex loop ``blowout''}, to be explained below. This
is confirmed by a detailed calculation of the distribution function for
loops of a given perimeter, as a function of the perimeter, for various
temperatures. In the low temperature regime, this distribution function
is an exponentially decreasing function of the perimeter, indicating that
there exists a length scale in the problem
associated with a typical size of thermally induced {\it closed}
vortex-loops in the system. However, in zero magnetic field there exists a
temperature scale, which we denote as $T_c$, at which the distribution
function is an algebraically decreasing function of the perimeter of vortex
loops, indicating that there no longer exists a length scale associated with
typical sizes of thermally induced closed vortex-loops. In this case, such
vortex-loops exists on {\it all} length scales in the problem, up to and
including the system size.\\
This means that the system experiences a thermally induced proliferation of
unbounded closed vortex-loops, a situation for which Onsager
coined the term ``vortex-loop blowout'' \cite{Onsager:Feynman}.
In zero magnetic field,
such a blowout marks the transition from a normal metal to a
superconducting state, or vice versa \cite{Tesanovic:B95,Dasgupta:L81}.
Many years ago, the vortex-loop blowout transition was suggested
to occur in the {\it neutral} super-fluid $He^4$ at the $\lambda$-transition
\cite{Onsager:Feynman}. The loop-transition in $He^4$ has more recently been
reinvestigated by several authors \cite{Williams:L87}. In the context of
{\it charged} super-fluids in {\it zero magnetic field} a corresponding
loop-transition was suggested to occur in {\it isotropic} lattice
superconductor models several years ago by Dasgupta and Halperin
\cite{Dasgupta:L81}, and more recently by Korshunov \cite{Korshunov:E90}.
The suggestion that features of this zero-field transition may survive
in finite magnetic fields, and thus be of importance for the statistical
mechanics of the {\it vortex-liquid phase}, has been suggested by
Te{\v s}anovi{\'c} \cite{Tesanovic:B95}, and considered recently by us
in detailed Monte-Carlo simulations of the Lattice London Model
\cite{Nguyen:L96}.\\
The main purpose of the present paper is to study, via detailed Monte-Carlo
simulations, the fate of the zero-field Onsager-Dasgupta-Halperin transition
when a magnetic field is applied to an extreme type-II superconductor,
using the somewhat simpler more familiar uniformly
frustrated anisotropic Villain-Model, which is related to the Lattice
London Model via a duality transformation. \\
For finite fields $B \neq 0$, we find two sharp features in the specific
heat and helicity modulus. In addition, we observe what appears to be a
cross-over at a considerably larger temperature, in agreement with recent
simulations \cite{Teitel:C96,Teitel:B93}. \\
The first sharp feature we find, at a temperature which we denote as $T_m$,
is identified as the first order melting transition of the FLL. The second
sharp feature is more subtle. It takes place at a temperature which we
denote as $T_z$. At the temperature $T_z$, we find that the phase-coherence
in the BCS superconducting order-parameter across the sample in the
direction of the magnetic field is destroyed. We also find, via our
computations, that flux-line cutting and the amount of intersecting flux
lines dramatically increases at $T_z$. Consequently, above $T_z$, phase
coherence along the field direction is destroyed.
{\it Furthermore, for the anisotropies considered in this paper, it occures
that $T_z \to T_m$ from above as the system size is increased.}
$T_z$ never drops below $T_m$, for reasons to be explained below.
We emphasize that we at this stage are limiting this statement to the case
of {\it infinite} penetration depth, since the results are obtained
within the uniformly frustrated Villain-model only. \\
{}From this we draw two conclusions. Firstly, when the flux-line lattice melts,
it does not melt into a flux-line liquid which has phase-coherence
along the direction of the applied magnetic field in any temperature
regime. This however does not
mean that the flux-line liquid is an entangled vortex system:
At $T_z \to T_m$ we find that flux-line cutting and intersectioning of
flux lines with closed vortex-loops of diameter on the scale of the
magnetic length, increases abruptly. Hence, the flux lines cannot be
considered as well defined in  the liquid phase.
Secondly, in the present simulations on the anisotropic, uniformly
frustrated Villain-Model, we never observe an entangled flux-line
{\it lattice phase} of the type described
by Frey {\em et al.} \cite{Frey:94} for the layered case. However, we
cannot access extreme anisotropies in our simulations, for reasons to
be explained in Section II.D, and therefore do not rule out the existence
of a ``supersolid phase" such as proposed in Ref. \cite{Frey:94}.\\
We also observe a crossover that takes place at a temperature which
marks the  onset of  strong diamagnetism, not associated with global
phase-coherence in the superconducting BCS order parameter, but with
phase-coherence throughout finite domains. It takes place at a temperature
well above both $T_m$ and $T_z$, which we denote as $T_{Bc2}$. Usually,
this cross-over is identified  with $B_{c2}$, the upper critical field,
and signals the onset of strong diamagnetic fluctuations. This crossover
is the remnant of the zero-field second order phase-transition that
marks the onset of the transition form metallic to superconducting
behavior. The filling fractions we consider in this paper, $f=1/32$ and
$f=1/72$,
may be converted into a magnetic field of the order of $10$ Tesla,
which is not particularly low. The present paper, therefore, does not
address the issue of how the crossover at $B_{c2}$ in finite large fields
evolves into the sharp second-order transition in zero field. This issue is
of fundamental importance, and remains open. \\
The rest of this paper is organized as follows. In Section II we describe the
uniformly frustrated anisotropic Villain Model along with the approximations
inherent in this description of a superconductor. We also give the connection
between these models and the Lattice London Model. Then we define the physical
quantities to be considered, and their measurements in the simulations. In
Sections III and IV we show and discuss our results for the zero-field case
and the finite-field case, respectively. Finally in Section V we summarize
our main findings. The derivations of the helicity moduli we consider,
both in terms of phase-variables and in terms of vorticities, are given in
two appendices.

\section{The Model and Definitions}
\subsection{The Model}
Our starting point is the anisotropic lattice superconductor model (LSM)
(which semantically should be distinguished from the Lattice London Model)
\cite{Dasgupta:L81,Korshunov:E90,Carneiro:B92}, defined by the partition
function

\bleq
\begin{eqnarray}
	Z & = &  \prod_{\vec{r}} \prod_{\nu= x, y,z}
	     \left( \int_{-\pi}^{\pi} \frac{d\theta}{2 \pi}
	     \sum_{m_\nu = -\infty}^{\infty}
	     \int_{-\infty}^{\infty} dA_\nu \right)
	     ~~ \exp(-H_{LSM}/k_B T) ~,
\nonumber  \\
	H_{LSM} & = & \frac{J_0}{2} \sum_{\vec{r}} \sum_{\mu = x,y,z} \left[~
	    \alpha_\mu ( \nabla_\mu \theta - 2\pi m_\mu - A_\mu)^2
	    +\frac{\lambda_\mu^2}{d^2} (\vec{\nabla} \times \vec{A})_\mu^2
            ~\right].
\label{Partition:LSM}
\end{eqnarray}
\eleq

Here, $J_{0}$ is the energy scale for the system. Furthermore,
$\alpha_\mu$ is the anisotropy parameter along the $\mu$-direction,
and $\vec{\nabla}$ denotes a lattice derivative.  The variable
$\theta(\vec{r}) \!\in\![-\pi,\pi)$ is the phase of the complex
superconducting order parameter $\Psi(\vec{r})$ at site $\vec{r}$ of a
three dimensional numerical lattice with lattice constant $d$,
$m_\mu(\vec{r})$ are integer variables defined on the
directed link between site $\vec{r}$ and site $\vec{r}+\hat{e}_\mu$,
where $\hat{e}_\mu$ is the primitive vectors for the cubic unit cell
($|\hat{e}_\mu| \keq d$, Fig.~\ref{GroundState}).
The contribution $A_\mu(\vec{r})$ to the gauge-invariant phase of the
order-parameter is related to the vector potential $\vec{A}_{vp}(\vec{r})$ by
\begin{eqnarray*}
	A_\mu(\vec{r}) \equiv \frac{2\pi}{\Phi_0}
        \int_{\vec{r}}^{\vec{r} + \hat{e}_\mu}
	d\vec{r}' \cdot \vec{A}_{vp}(\vec{r}') ~,
\end{eqnarray*}
where $\Phi_0=2.07 \cdot 10^{-15} T m^2$ is the flux quantum. Finally,
$\lambda_\mu$ is the London penetration depth along the $\mu$-direction.\\
In this model, we neglect fluctuations of the amplitude in the
complex superconducting order parameter, i.e.
$\Psi(\vec{r}) \keq |\Psi(\vec{r})| e^{i\theta(\vec{r})}
\ksim \Psi_0 e^{i\theta(\vec{r})}$.
The Lattice London Model is obtainable from the Lattice Superconductor
Model by explicitly performing the $\theta$ and $A_{\mu}$-integrations
in Eq. (\ref{Partition:LSM}), as shown first by Korshunov and more
recently by Carneiro \cite{Korshunov:E90,Carneiro:B92}. \\
To study the physics of high-$T_c$ superconductors, we consider a three
dimensional {\em cubic lattice, with linear dimension $Ld$}, and with a
uniaxial anisotropy $\Gamma \!\equiv\! \lambda_c / \lambda_a$. In these
simulations the applied
magnetic field, and hence the net magnetic induction $B$, is taken along the
crystal $\hat c$-axis. Here, $\lambda_a$ and $\lambda_c$ are the penetration
depths in the crystal ab-plane ($CuO_2$-plane) and along the crystal
$\hat{c}$-axis, respectively. Subsequently, we will take the limit
$\lambda_a,\lambda_c \to \infty$, but in such a way that the ratio
$\lambda_c/\lambda_a$ is maintained constant. We take our coordinate
$(\hat{x},\hat{y},\hat{z})$-axis parallel to the crystal
$(\hat{a},\hat{b},\hat{c})$-axis, respectively. Periodic boundary conditions
in all direction are assumed. The basic parameters of the LSM are given by
\cite{Carneiro:B92},
\begin{eqnarray*}
	J_{0} = \frac{\Phi_{0}^{2}d}{16\pi^{3}\lambda_{a}^{2}},
	~~~~~~~~~~
	\alpha_\mu = \frac{\lambda_a^2}{\lambda_\mu^2}.
\end{eqnarray*}
Here, $d$ may tentatively be interpreted as the distance between two
Cu$O_2$-layers {\it in adjacent unit cells}.
The vorticities $n_x(\vec{r}),n_y(\vec{r})$ (corresponding to fundamental
vortex-line segments parallel to the ab-plane, defined in Eq.~\ref{cond2})
are assumed to exist {\it in between} $CuO_2$ double-or multiple-layers
in compounds such as YBCO and BSCCO. We use the numerical lattice unit as
a measure of the in-plane coherence length,  $\xi_a \ksim d$. Note that
since the numerical factor relating $\xi_a$ to $d$ is not uniquely
determined in our approach, the filling fraction $f$ does not uniquely
determine the magnitude of the applied magnetic field. Variation of $f$ may
thus
be viewed as a variation of $B$, but alternatively a reduction of $f$
may be viewed as an improvement of the approximation to the continuum
limit at fixed induction $B$.) \\
As for the London Model, the LSM is appropriate for
describing the physics of extreme type-II
superconductors ($\lambda_a \gg \xi_a$) in the field regime $B \ll B_{c2}$,
where $B_{c2}$ is the upper critical magnetic field, implying
$a_v \gg \xi_a$. Thus, spatial variations of the {\it amplitude}
of the superconducting order parameter may be neglected. In these
simulations it is also postulated that the penetration length is
essentially infinite, which in practice means that they are at least
much larger than the average distance between vortex lines, when the field
is finite. Hence we also have the requirement $B \gg B_{c1}$. In terms of
magnetic induction, our simulations are thus strictly speaking limited to
the field-regime $B_{c1} \ll B \ll B_{c2}$ {\it when finite fields are
considered}. For our zero-field results, the complete suppression of
gauge-fluctuations implies that the penetration depth of the model must be
at least larger than any system-sizes considered. \\
The Monte Carlo simulation time $t_{MC}$ for the LSM on a cubic system
with linear dimension $L$ is of order $L^6$. The suppression of the
gauge-field fluctuations, using the  limit $\lambda_\mu$=$\infty$, reduces
the required computer time dramatically, $t_{MC}$ to $\ksim L^3$. The neglect
of gauge-fluctuations reduces the  LSM to the uniformly frustrated
anisotropic Villain Model \cite{Williams:L87,Teitel:B93,Villain:Paris75},
which is the model used in this paper. It is defined by  the following
partition function after performing the sum over $m_\mu(\vec{r})$ in
Eq. \ref{Partition:LSM} explicitly,
\begin{eqnarray}
	Z & = & \prod_{\vec{r}}
	    \left ( \int_{-\pi}^{\pi} \frac{d\theta}{2 \pi} \right )
            ~~ \exp(-H_v /k_B T) ,
\nonumber  \\
	H_v & = & J_0 \sum_{\vec{r}} \sum_{\mu = x,y,z}
	    V_\mu(\nabla_\mu \theta - A_\mu;T),
\label{Partition:Villain} \\
	V_\mu(\chi; T) & = & -\frac{k_B T}{J_0}
	    ln \left \{ \sum_{m = -\infty}^{\infty}
	    \exp \left [-\frac{J_0 \alpha_\mu}{2k_B T}
            (\chi - 2 \pi m)^2 \right ] \right \}.
\nonumber
\end{eqnarray}
The advantage of the Villain-model model
compared to the Lattice London Model used in earlier large-scale simulations
we have performed on the Abrikosov vortex lattice \cite{Nguyen:L96}, is
that is allows considerably larger system sizes to be studied than with the
Lattice London Model. The latter model has the intuitively appealing feature
of allowing simulations on line-like objects, but as we have seen, the
Villain-model and the Lattice London model are in principle equivalent
representations of a lattice superconductor model. One other
major advantage of the Villain-model compared to the Lattice London
model is that, while it is straightforward to extract unambiguous
information about vorticities from the phases of the Villain-model,
it is impossible to reconstruct
unambiguous  phase-information from the vortex-degrees of freedom in
the Lattice London model. Thus the Villain-model straightforwardly
provides information on vorticities as well as phase-coherence.
Ultimately, the choice of model to use in simulations is dictated
exclusively by convenience, and depends to a large degree on what
problems to consider. One problem where the Lattice London Model appears
to present clear advantages over the Villain-model is the problem of
numerical simulations of flux-creep in the presence of pinning. \\
The uniformly frustrated anisotropic Villain Model is appropriate for
describing the physics of extreme type-II superconductors in the limit where
the penetration depth is larger or comparable to the system size (zero
magnetic field) or when the penetration depth is much larger than the
average distance between flux lines $\lambda \gg a_v$ (finite magnetic
fields).

\subsection{The Ground State}
\begin{figure}[htbp]
\psfig{figure=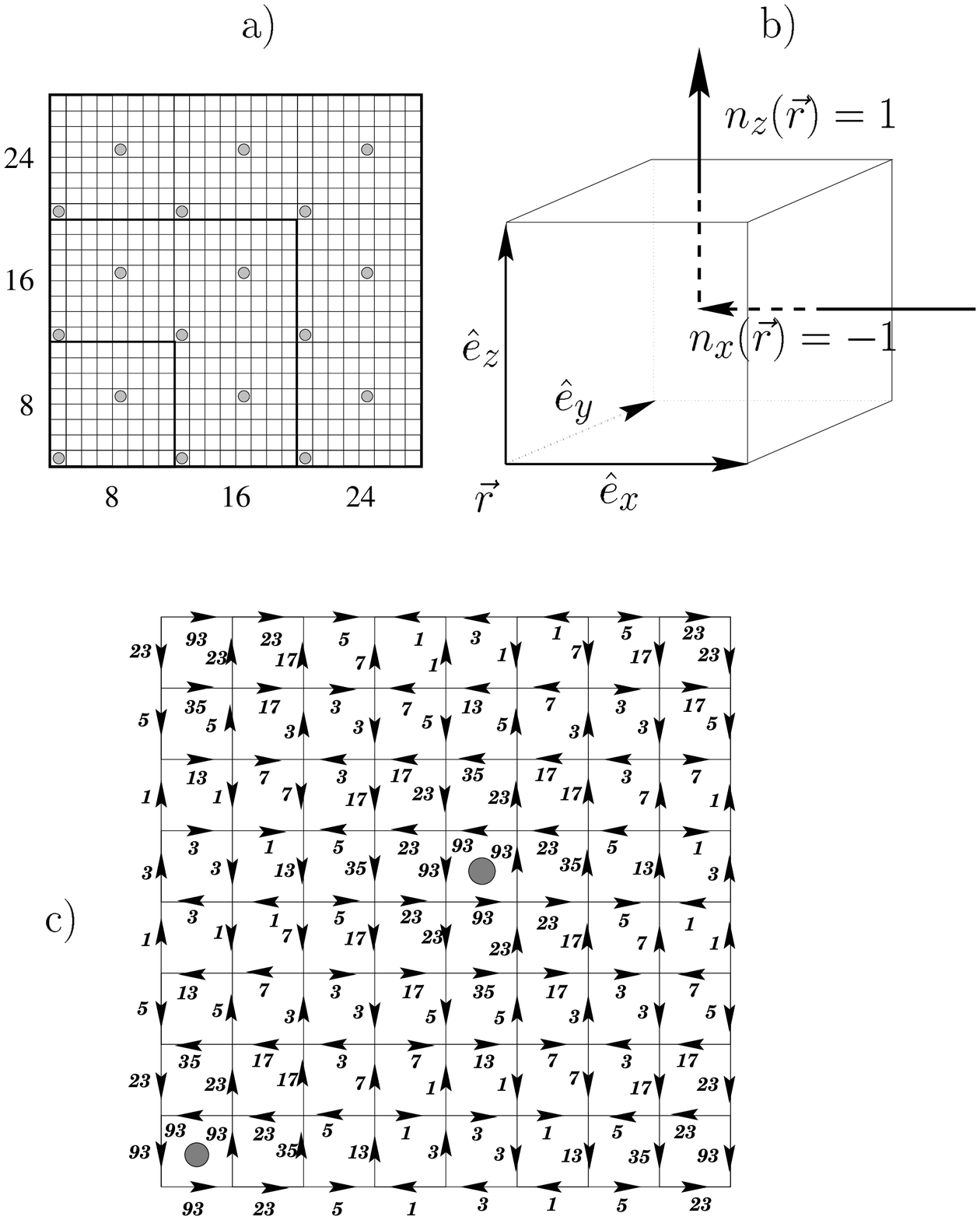,height=13cm,angle=0}
{\small FIG.~\ref{GroundState}
	     a) The ground state flux-line configuration
                for f=1/32 and system sizes L=8,16,24.
             b) A cubic unit cell with two elementary vortex segments
                penetrating two plaquettes of the unit cell.
	     c) The ground state current pattern in the ab-plane
		in units of $2\pi/384$ for the 8x8 vortex lattice
                unit cell. The arrows indicate the direction of the
                currents on each link. The current pattern in this ground
                state is complicated, but nevertheless exhibits a high
                degree of symmetry.}
\refstepcounter{figure}
\label{GroundState}
\end{figure}

The current corresponding to the gauge invariance phase differences
\begin{eqnarray*}
	j_\mu(\vec{r}) =   \theta(\vec{r} + \hat{e}_\mu) - \theta(\vec{r})
			 - A_\mu(\vec{r}),
\end{eqnarray*}
is defined on the directed link between site $\vec{r}$ and site
$\vec{r}+\hat{e}_\mu$, $j_\mu(\vec{r}) \!\in\! [-\pi,\pi)$.
This current obeys two conditions:
(1) There are no net current sinks or sources in the ground state
	\begin{equation}
		\sum_{\nu=x,y,z}\nabla_\nu j_\nu(\vec{r}) \keq 0.
	\label{cond1}
	\end{equation}
(2) The counterclockwise line integral  of the currents around any
plaquette of the numerical lattice with a directed surface normal
in the $\mu$-direction at site $\vec{r}$ {\em must always} be
	\begin{equation}
		\sum_{Ci} j_\nu(\vec{r}) =
			2\pi(n_\mu(\vec{r}) \km f_\mu).
	\label{cond2}
	\end{equation}
Here, $Ci$ is the closed path traced out by the links surrounding an
arbitrary  plaquette,
and $\nu$ represents the Cartesian components of the current
in the directions of the links which comprise the closed path $Ci$.
Furthermore, $n_\mu(\vec{r}) \keq 0$, $\pm 1$ represents a vortex
segment {\it penetrating the plaquette enclosed by the path $Ci$}.
The situation is illustrated in Fig.~\ref{GroundState}.
Furthermore, $f_\mu$ is the filling fraction along the $\hat{\mu}$-direction,
defined in Eq.~\ref{fmu}.
In this way, we can find {\em the distribution of vortex segments
$\vec{n}(\vec{r})$}, by calculating the counterclockwise line integral of
the currents around every plaquette in the system.  Hence, the distribution
of gauge-invariant current also gives information, essentially by a duality
transformation, about the FLL structure function.

To perform a finite size scaling analysis, we employ the following procedure
to find the current pattern in the ground state. Given a density of
flux lines $f \keq 1/32$, we design an $8 \times 8$ {\it vortex lattice unit
cell}, not to be confused with the unit cell of the numerical lattice. This
vortex lattice unit cell has two vortices, Fig.~\ref{GroundState}.
The current pattern (Fig.~\ref{GroundState}) is found by requiring the
currents to obey Eq.~\ref{cond1} on every link and Eq.~\ref{cond2} on every
plaquette throughout the vortex lattice unit cell.
It is possible to reduce the number
of unknown currents by requiring the current pattern to have the same
symmetry as the ground state vortex lattice. Periodic boundary conditions at
the boundaries of the vortex lattice unit cell are used. \\
By repeating the vortex lattice unit cell, we can design the current
pattern of all systems with size m8$\times$n8$\times$l, (m,n,l) being
positive integers.

The flux-line density  along the $\mu$-direction is defined as
$f_{\mu}$, and is given by
\begin{eqnarray}
	f_{\mu} = \frac{\sum_{\vec{r}}n_{\mu}(\vec{r})}{L^3}.
\label{fmu}
\end{eqnarray}
In the ground state, a uniform magnetic induction along the crystal
$\hat{c}$-axis $\vec{B} = B \hat{z}$ gives a periodic structure of
straight flux lines aligned with $\vec{B}$ with hexagonal symmetry
on a continuum substrate ab-plane, the well known hexagonal
Abrikosov vortex lattice.  In terms of the above densities
$f_{\mu}$ this is expressed as
\begin{eqnarray*}
	f_z = \frac{B d^2}{\Phi_0} \equiv f, ~~~
	f_x = f_y = 0.
\end{eqnarray*}

In our simulations, it is not possible to exactly load the hexagonal
Abrikosov vortex-lattice onto our numerical mesh, which we have chosen
to be square. This means that the
underlying numerical mesh necessarily introduces a distortion of the
hexagonal ground state. The numerical mesh represents a commensuration
potential which acts as a perturbation on the ground state, and tends
to `freeze' the flux lines into a structure commensurate with it. The flux
lines will however tear themselves off such a commensuration potential
caused by the numerical lattice
at a high enough temperature, which we denote a `depinning' temperature
$T_{dp}$. Note that this depinning temperature has nothing to do
with a real pinning potential, it is purely an artifact of the
underlying numerical lattice. In the continuum limit, it would be zero. It
is at present unclear to what extent the numerical mesh represents a
singular perturbation on the continuum limit in a $3D$ system. The well
known results of Nelson and Halperin \cite{NH:L78} and Young \cite{Young:B79}
concerning the effects of periodic commensuration potentials in $2D$,
indicates that if the filling fraction is small enough in $2D$, $T_{dp}$ will
be smaller than any other relevant energy scale in the problem. This has
also been nicely confirmed in a number of recent simulations on $2D$ systems
\cite{Franz:L94}. \\
Note that, by using a square numerical mesh, we counteract the disadvantage
of the distortion of the hexagonal lattice by a reduction in the strong
commensuration effects we would have encountered if we had chosen a
triangular numerical mesh, which admits an exact hexagonal lattice ground
state. \\
By using low enough filling fractions $f_z$, it may be  hoped even
in $3D$ to achieve a satisfactory approximation to the continuum limit. That
is, we hope that the depinning temperature $T_{dp}$, which appears purely
as an artifact of defining the model on a numerical lattice, drops below
all other relevant temperatures in the
problem, including the putative melting temperature $T_m$ of the FLL. \\
That such a thermal depinning from the numerical lattice can actually
be achieved in higher dimensions  than $2D$ is by no means clear, since
commensuration effects are much more pronounced in $3D$ than in $2D$. In
fact, in the thermodynamic limit, thermal depinning from the numerical
lattice strictly speaking {\it cannot} happen in $3D$. Therefore, in order
to mimic the statistical mechanics of the FLL defined on a continuum
substrate in $3D$, one carefully has to choose sample-geometry in
simulations as follows: It is crucial to have a sample-geometry where the
distance between flux lines is tailored to suit the thickness of the
sample in such a way that the commensuration
potential along the direction of the flux lines does not  cause pinning
to the numerical lattice at the temperatures of interest. This implies that
simulations must be carried out on relatively flat slabs. Simulations
on `tower'-shaped slabs \cite{Teitel:B93} are probably not able to capture
the continuum limit. In our simulations, we have checked that in fact
such a depinning transition from the numerical mesh takes place
at a temperature $T_{dp}^{3D}$ which is below the temperatures
of primary interest. Hence, in our simulations  the continuum limit
ought to be adequately mimicked. Note that for filling fractions
$f=1/8$ and $f=1/15$, the continuum limit is not mimicked satisfactorily.
In these cases, it is clear that pinning to the numerical mesh strongly
influences the results.\\
In this paper, we consider mainly the case of $f \keq 0$ and $f \keq 1/32$,
while some results are also obtained for $f=1/72$.

\subsection{The Monte Carlo Simulation}
The statistical mechanics of the Villain-model of an extreme type-II
superconductor is investigated by employing the following Monte Carlo
procedure on numerical cubic lattices with linear dimensions
$L~\in~[8,16,24,32,40,48,64,80,96]$. \\
Identical sets of current patterns are loaded onto each layer of the
numerical lattice. For the filling fraction $f=1/32$, this current pattern
is illustrated in Fig. \ref{GroundState}. We update the system, heating
the system  from the ground state consisting of $fL^2$ straight
field-induced flux lines. A site of the numerical lattice is chosen
randomly, and an attempt is made to change the phase on that site with a
random amount $\Delta \theta \in [-\pi, \pi)$. The phase change is accepted
or rejected according to the standard Metropolis algorithm. \\
If the accepted phase change causes the current on a link $j_\mu(\vec{r})$
to fall outside the range  $j_{\mu}(\vec{r}) \in [-\pi, \pi)$ we add an
amount $\pm 2 \pi$ to the current, such that $j_\mu(\vec{r})$ is brought
back into the
primary interval $j_{\mu}(\vec{r}) \in [-\pi, \pi)$. An important
point is that this operation can only generate a closed unit vortex loop
around the link where the current is changed, thereby conserving the net
induction of the system. No net vorticity is ever introduced by the
procedure, and the procedure also guarantees that no  flux line can start
or end within the sample. It is also important to note that the Monte Carlo
procedure described above satisfies detailed balance. Hence, the entire
phase space of the Villain-model is guaranteed to be exhausted provided
the simulations are run for a long enough time. Another point is that the
above procedure for limiting the currents to the primary interval also
limits the number of vortex segments penetrating a plaquette to at most
one per plaquette. In this way, the Villain-model differs from the Lattice
London Model, where the number of vortices penetrating  each plaquette can
be arbitrary. This difference however only becomes important in the
high-temperature regime.\\
The Monte-Carlo procedure really updates the gauge-invariant
phase-differences, or currents $j_{\mu}(\vec{r})$. The simulations are
therefore carried out in a manifestly gauge-invariant manner. One Monte
Carlo sweep consists of $L^3$ attempts to change the  phases
$\theta(\vec{r})$
on $L^3$ randomly chosen sites throughout the lattice. Thus, by such a move
we simultaneously change the gauge-invariant phase-differences on the six
links associated with the relevant lattice point. Each data point for the
quantities we consider is obtained after discarding the first $100000$
$(30000)$ sweeps for equilibration. The subsequent $400000 ~ (70000)$
sweeps are used to obtain averages. The numbers in parentheses represent
the number of sweeps we have for the case of $\Gamma = 1$.
To ensure that measurements are independent of each other, we
do one measurement per 100 sweeps.

\subsection{Anisotropy and finite size-effects}
For $B \keq 0$ and isotropic couplings, the LSM, the XY Model, and the
Villain Model all have one single phase transition at
$k_B T_c/J_0 \ksimeq 3.0$. The transition is characterized by stiffness in
the phase of the superconducting order parameter being lost across the
system in all directions, due to a blowout of thermally excited closed
vortex loops. \\
For the anisotropic case, the bare coupling between planes $J_\perp$ in
the Villain Model is smaller than the in-plane coupling $J_\parallel$.
Thus, in the very anisotropic case, the excitation energy of a unit vortex
loop parallel to the ab-plane is much smaller than the excitation energy
for a unit vortex loop containing segments perpendicular to the ab-plane
\cite{AS:L91}. One would naively then expect that thermal excitation of
vortex-loops parallel to the ab-planes would occur at correspondingly lower
temperatures than those for which vertical loops appear. This is true for
unit vortex loops, but such loops are unimportant for critical behavior.
{}From the point of view of considering $B=0$ critical phenomena, the important
issue is how the anisotropy affects large vortex loops, including vortex
loops of order the system size. This is an issue to which we now briefly
turn. It is convenient to carry out this discussion in terms of the phases
of the superconducting order parameter, rather than in terms of
vorticities.\\
When the temperature approaches the Kosterlitz-Thouless (KT) temperature
from above in a quasi-$2D$ system, the phase-coherence length gradually
grows. In a strictly $2D$ system, it would diverge precisely at the
KT-transition. However, as long as a small coupling between planes
exists, no matter how small, then as the KT-transition is approached from
above, increasingly larger domains of correlated phases are coupled together
by the inter-plane coupling. This strongly renormalizes the bare inter-plane
coupling constant $J_{\perp}$ \cite{Korshunov:E90,Horovitz:B93}. Hence, the
system is isotropized close to, but above, the KT-transition, and
the transition retains a $3D$ character. Thus, even an extremely
anisotropic  system exhibits, precisely as in the isotropic case, one
single $3D$ phase transition. No decoupling transition as proposed in
Ref. \cite{Friedel:JP88} exists in zero magnetic field. \\
In simulations on finite systems, care must be taken to ensure that
this physics is captured correctly. When the anisotropy of the system is
increased such that the bare interplane coupling is reduced, one must make
sure that the dimensions of the system in the directions parallel to the
ab-planes are large enough to allow the required renormalization of the
coupling constants to run its course without being cut off prematurely by
the system size.  Equivalently, one may rescale the size of the system
in the $z$-direction, by the factor $1/\Gamma$. Thus, the transverse system
size must be tailored to the anisotropy of the system in such a way that
critical behavior we study takes place at a {\it lower} temperature than
the energy scale $k_B T^*$ set by
\begin{eqnarray*}
k_B T^* = \left ( \frac{\xi}{d} \right )^2 ~~ J_{\perp}.
\end{eqnarray*}
Here, $J_{\perp}$ is the bare inter-plane coupling in the Villain-model,
and $\xi$ is the coherence length of the phase of the superconducting order
parameter at the relevant temperature. If the system is too small in the
transverse direction, the renormalization of the bare coupling is cut
off by the system size
\begin{eqnarray*}
k_B T^* = L^2 ~~ J_{\perp}.
\end{eqnarray*}
Hence, for a given $L$, we may choose an anisotropy such that $J_{\perp}$
is so small that $T^*$ becomes smaller than the actual temperature of
the $3D$ critical phenomenon of interest, namely the vortex-loop
blowout. We would then observe a decoupling of planes due to a proliferation
of vortex-loops in the ab-plane which would be unphysical.  \\
This finite-size effect limits the anisotropies we can
study consistently, at least in zero magnetic field. In zero field, we
find ourselves limited to anisotropies of $\Gamma \alt 4$. \\
A final technical point is that, although a finite magnetic field
{\it a priori} allows larger anisotropies to be studied, the
Villain-potential itself becomes virtually featureless as a function of
its arguments $\chi$, Eq. \ref{Partition:Villain}, for large anisotropies.
Hence, simulations become impossible to perform meaningfully.

\subsection{The helicity modulus}
To probe the global phase coherence in the BCS superconducting order
parameter across the entire system, we consider the helicity modulus
$\Upsilon_\mu$, defined as the second derivative of the free energy with
respect to a phase twist in the $\mu$-direction \cite{Teitel:B93}.  It
basically measures the stiffness of the system to a twist in
the phase of the order parameter. In the anisotropic case we have the
following generalization of previously obtained expressions for the helicity
modulus for isotropic superconductors \cite{Teitel:B93}
{\small
\begin{eqnarray}
	   & \Upsilon_\mu &  =
  	   \frac{J_0^2}{L^3 k_B T} \left \langle \sum_{\vec{r}, \nu}
           V'_\nu [~ \theta(\vec{r} + \hat{e}_\nu) - \theta(\vec{r})
	   - A_\nu(\vec{r}) ~] (\hat{e}_\nu \cdot \hat{e}_\mu) \right\rangle^2
\nonumber   \\
	   & - & \frac{J_0^2}{L^3 k_B T} \left \langle \left (
	   \sum_{\vec{r}, \nu}
           V'_\nu [~ \theta(\vec{r} + \hat{e}_\nu) - \theta(\vec{r})
	   - A_\nu(\vec{r}) ~] (\hat{e}_\nu \cdot \hat{e}_\mu)
           \right )^{\! \! \! 2} \right \rangle
\nonumber  \\
 	   & + & \frac{J_0}{L^3} \left \langle  \sum_{\vec{r}, \nu}
           V''_\nu [~ \theta(\vec{r} + \hat{e}_\nu) - \theta(\vec{r})
           - A_\nu(\vec{r}) ~] (\hat{e}_\nu \cdot \hat{e}_\mu)^2
	   \right\rangle . \!\! \!
\label{helmod}
\end{eqnarray} }
For the details on the derivation of this expression, and a corresponding
one in terms of vorticities, see Appendices A and B. Here, $V'_\mu$ and
$V''_\mu$ are the first and second derivatives, respectively, of the
anisotropic Villain potential $V_\mu$ defined in Eq.~\ref{Partition:Villain}.
For temperatures $T < T_{\mu}$ such that $\Upsilon_\mu \kgt 0$, there is
phase coherence across the entire system in the $\mu$-direction. Hence, the
system can sustain a supercurrent in the $\mu$-direction. At the temperature
$T=T_{\mu}$ and above, such that $\Upsilon_\mu \keq 0$, phase-coherence along
the $\mu$-direction is lost. Hence, the vanishing of $\Upsilon_\mu$ signals
the superconducting normal metal transition in the $\mu$-direction with the
transition temperature $T_\mu$. For the case of finite magnetic induction
along $\hat{z}$-axis in the continuum limit, $\Upsilon_x$ and $\Upsilon_y$
should vanish at all temperatures \cite{Teitel:L94}, since a current in the
ab-plane would exert a Lorentz-force on the unpinned flux lines, moving them
and thus dissipate energy. When the model is discretized by introducing the
numerical lattice, a finite energy for moving them in a direction perpendicular
to the $\hat z$-direction is also introduced. The existence of a smallest
energy to required move flux-lines acts as an artificial pinning potential
on the FLL, causing $T_x~(T_y \keq T_x)$
to be finite. Thus, the FLL 'depins' from the underlying discrete lattice at a
finite temperature $T_x = T_y > 0$. In the continuum limit, as long as no
physical pinning of the flux lines is present, we would have $T_x=T_y=0$,
and the flux lines are unpinned at all temperatures. To ensure that the
above artificially introduced  pinning potential caused by the numerical
lattice does not affect the FLL melting transition and any other genuine
phase-transition we might want to consider, we should consider systems with
$T_m$ significantly higher than $T_x$. The way to achieve this is to
consider low enough filling fractions $f=f_z$ of flux lines. Several authors
\cite{Nguyen:L96,Franz:L94} have in fact found that $T_x$ decreases for
decreasing flux lines density f, and falls below $T_m$ for
$f \klt f_c \ksimeq 1/32$. \\

\subsection{Vortex-Loop Distribution}
As mentioned in the Introduction, in the LSM, Villain-model, and Lattice
London Model, the zero-field normal metal-superconductor transition
corresponds to a vortex-loop blowout analogous to what has long been
suggested to occur in the neutral superfluid $He^4$. To study the
blowout of closed vortex loop in extreme type-II superconductors, we
consider the quantity $D(p)$, which we denote the vortex-loop distribution
function, and which is given as statistical average of the total number
of closed {\it non-field induced} vortex loops with a given perimeter
$p$, in our case normalized by the volume of the systems we consider. \\
The following procedure is employed to compute $D(p)$. \\
We start from an arbitrarily chosen unit cell containing at least one
{\it outgoing} vortex segment penetrating a plaquette of that unit cell. We
then follow the  direction of this vortex segment {\it into} the neighboring
unit cell. If there is more than one vortex segment leaving the unit cell,
one of them is chosen randomly. We continue tracing the path of vortex
segments until the path closes up on it self. When the path
has closed upon itself, we measure the length
$l$ of the path, as well as the net vorticity $v_z$ along $\hat{z}$-axis
of the path. Also, we remove the vortex segments along the path to  prevent
double counting of paths. Because of the periodic boundary conditions,
such closed paths of vortex segments can either belong to a field-induced
flux line or a non field induced closed vortex loop. \\
In the quantity $D(p)$, we do {\it not} include the closed paths associated
with field-induced flux lines that close on themselves merely due to periodic
boundary conditions in the $\hat z$-direction. Field-induced flux-lines are
characterized by a net vorticity in the $\hat z$-direction, $v_z \neq 0$.
For the purposes of studying the loop-transition, we are exclusively
interested in closed paths associated with {\it non-field induced}
vortex loops that physically close on themselves regardless of
boundary conditions. Hence, the  relevant closed paths of vortex segments
are closed vortex loops with perimeter $p \keq l$ which have net
vorticity $v_{\mu}=0$ in all directions. \\
The procedure for tracing out the relevant closed loops is repeated until
all non-field vortex segments in the system have been counted. For
$B \not = 0$, this procedure uniquely separates thermally excited closed
vortex loops from the field induced flux lines. \\
In the low temperature regime $D(p)$ depends on the excitation energy
$E(p) \sim \varepsilon p$ of the vortex loops with perimeter $p$,
with a Boltzmann-factor \cite{Teitel:B93},
\begin{eqnarray*}
	D(p) \ksim \exp(- \frac{\varepsilon p}{k_B T})
	     \ksim \exp(- \frac{p}{L_0}),
\end{eqnarray*}
where $\varepsilon$ is a constant representing a line tension, and
$L_0 \sim k_B T/\varepsilon$ is a typical perimeter of closed vortex-loops
present at a given temperature {\it in the low-temperature regime}. As we
will see below, such low-temperature ``confined"
vortex-loops may be
coarse-grained away and are unimportant for the statistical mechanics of
the mixed state of an extreme type-II superconductor. In this thermally
activated regime, large vortex loops are exponentially suppressed. On the
other hand, at the critical point, vortex loops with all perimeters are
present. This leads to an algebraic decay of the loop-distribution function
versus loop-perimeter at the critical point, \\
$D(p)~\ksim~p^{-\alpha}$,
where $\alpha$ is an exponent not to be confused with the critical exponent
of the specific heat. Hence, monitoring the temperature where $D(p)$
changes its characteristic behavior from exponential decay to algebraic
decay, is a way of determining the vortex-loop unbinding temperature.\\
If we assume that the vortex-loop-distribution function scales with the
vortex-loop perimeter as some power-law, and furthermore assume that
the perimeter scales with the vortex-loop radius $r$, then we have
\begin{eqnarray*}
D(r) \sim r^{-\alpha}.
\end{eqnarray*}
We now use a critical scaling analysis to
determine $\alpha$ in our case. The assumption is that the loop-transition
in zero field represents a critical point. If we can then fit numerically
obtained exponents at the putative critical point to the scaling results,
this would provide further support for the assertion that the loop-blowout
is responsible for destroying superconductivity in extreme type-II
superconductors. \\
The Villain-model is dual to a $3D$ Coulomb-gas, and the vortex-loops are
analogous to the vortex-antivortex pairs of the $2D$ Coulomb-gas. We may
determine the contribution to the dielectric constant of the $3D$
Coulomb-gas that such loops give. Since the dielectric constant may be
related to the inverse superfluid stiffness, whose scaling dimension is well
known on general grounds, we may determine $\alpha$ at the critical point.  \\
The ``dipole-moment'' $P(r)$ of a vortex-loop  scales with $r^{d-2} \cdot r$,
where the charge $\sim r^{d-2}$, and the dipole vector $\sim r$. The
contribution to the dielectric constant, or the electric susceptibility,
coming from thermally induced loops of size between $r$ is given
by \cite{KT:73}
\begin{eqnarray*}
   \chi_e (r)\sim \varepsilon(r) \sim \frac{\partial}{\partial E}
   <r^{d-1} \cos(\phi)>_{|{E=0}},
\end{eqnarray*}
where the average should be a thermal average with the Boltzmann-factor
\begin{eqnarray*}
\exp{[- U(r)/k_B T]} \sim D(r) ~~ \exp{ [r^{d-1} ~ E ~ \cos(\phi)/k_B T] }.
\end{eqnarray*}
Here $E$ is an electric field polarizing the medium via the ``charge-loops''
of the $3D$ Coulomb-gas, and $\phi$ is the angle between the orientation of
loops and the applied electric field polarizing the medium. We find
\begin{eqnarray*}
\varepsilon(r) \sim r^{2(d-1)-\alpha}.
\end{eqnarray*}
On the other hand, in the superconductor, the superfluid stiffness $\rho_s$
is given by the transverse susceptibility $\chi^{\perp} \sim G \sim \rho_s$,
where $G$ is given by the order-parameter Green's function,
$G \sim r^{2-d-\eta}$, and where $\eta$ is the anomalous scaling dimension
of the Green's function appearing due to critical fluctuations, for the
$3D XY$-model we have $\eta = 0.033(4)$. Now we use
the fact that $\varepsilon \sim \rho_s^{-1}$, to find
\begin{eqnarray*}
2(d-1) - \alpha & = & d + \eta - 2, \\
\alpha  & =  & d - \eta.
\end{eqnarray*}
In our case, we may evaluate the loop-distribution function at the anomalous
peak in the specific heat, and fit the result to a power law with the
exponent $3$, obtained by assuming critical scaling. We will see below that
the fit is excellent, lending further support to the assertion  that in zero
field, the superconductor-normal metal phase-transition in an extreme
type-II superconductor is due to a vortex-loop blowout transition.  \\
It is also interesting to note that another way of estimating the relevance
of closed vortex-loops, is to see whether they can be coarse-grained away or
not. A rough criterion for coarse-graining them away, would be that the
loop-distribution integrated over the volume of the system should be finite,
i.e.
\begin{eqnarray*}
   \int d^dr ~ D(r)  \sim \int_0^{\Lambda} ~ dr ~ r^{d-1-\alpha} \sim
   \Lambda^{d-\alpha}.
\end{eqnarray*}
Thus, loops may be coarse-grained away provided that $\alpha > d$.
In the low-temperature regime, we have seen that the loops are even
exponentially suppressed, and certainly satisfy this criterion. The
marginal case $\alpha=d$ gives an integrated distribution
$\sim \ln \Lambda$. However, at the critical point,
$\alpha = d - \eta$ is less than the required value for coarse-graining,
furthermore critical fluctuations as manifested by a non-zero
positive value of the exponent $\eta$, will increase the relevance
of vortex-loops, as expected.

\subsection{Specific heat}
In addition to indirect measurements of the latent heat of melting,
such as measured by local magnetization measurements on BSCCO
\cite{Zeldov:N95}, direct calorimetric measurements of the specific
heat are also useful for estimating the latent heat of the melting
transition of the FLL, or any other phase-transition the vortex system
might suffer.
Such measurements are now available \cite{Schilling:N96} both in zero field
and in finite field. In \cite{Schilling:B95}, the specific heat anomaly
in a twinned YBCO sample is measured systematically with varying magnetic
field. It evolves smoothly from the zero field result as the magnetic
field is increased. Moreover, the integrated anomaly appears to
be approximately constant as the magnetic field increases. \\
This raises the question of what sort of phase-transition, if any, the
specific heat anomaly in finite magnetic fields should be associated with.
Due to its smooth evolution from the zero-field case, it appears rather
unlikely that this sizeable  anomaly has anything whatsoever to do with FLL
melting.  Rather, it seems to suggest that there are remnants of the
zero-field transition, which we will describe in detail below, at finite
magnetic fields. \\
We may investigate this issue, by calculating the specific heat of the
Villain model, and correlate the specific heat anomalies with the
temperature dependence of the structure function $S({\bf K})$ as well as
with  the phase-stiffness $\Upsilon_z$. Thus we should in principle be able
to decide whether or not the major features in the specific heat have
anything to do with FLL melting or vortex-loop blowout, also at finite
fields. In zero field, we will be able to precisely
correlate, for all anisotropies, the anomaly in the specific heat with the
vortex-loop blowout transition. In a finite field, the situation is
considerably more complicated. We find three anomalous features in the
specific heat. The major feature in
the specific heat, the remnant of the zero-field transition, occurs at
temperatures well above those where the structure function of the FLL, and
the phase-stiffness across the sample along the field direction,
vanishes. \\
To calculate the specific heat per site $C$, we use the usual fluctuation
formula
\begin{eqnarray*}
   \frac{C}{k_B} = \frac{1}{L^3} \frac{<H_v^2> - <H_v>^2}{(k_BT)^2}.
\end{eqnarray*}
The Villain model has a rather unusual property in that the Boltzmann
factor appearing in the partition function, $\exp(- H/k_B T)$ involves an
explicitly temperature dependent Hamiltonian, Eq. (\ref{Partition:Villain}).
The usual fluctuation formula for the specific heat is valid strictly
speaking only if $H$ is temperature
independent. Nevertheless, we  will use the above standard expression for
calculating the specific heat for convenience, and neglect the extra terms
that should be included from the explicit temperature dependence of the
Villain potential, Eq.~\ref{Partition:Villain}.
We have checked the validity of this approximation by comparing the thus
obtained specific heat per site $C$ with the alternative standard method of
extracting the specific heat per site $C$ from its basic definition in terms of
the internal energy of the system
\begin{eqnarray*}
   \frac{C}{k_B} = \frac{1}{L^3} ~ \frac{\partial <H_v>}{\partial (k_B T)}.
\end{eqnarray*}
We find that these two ways of calculating the specific heat
differs only in the very high temperature regime, outside the
temperature range of interest in  this paper.\\
\subsection{The Structure Function}
To probe the FLL melting, we consider the structure function
for $n_z$ vortex segments, i.e. vortex segments directed along
the average induction. The structure function $S({\vec{k}})$
is defined by  \cite{Nguyen:L96,Carneiro:E92}
\begin{eqnarray*}
  S(\vec{k}) = \frac{<\mid \sum_{\vec{r}} n_z(\vec{r})
                      \exp ~ [i \vec{k} \cdot \vec{r} ]\mid^2>}
	            {(fL^3)^2}.
\end{eqnarray*}
For our ground state with the flux lines density $f \keq 1/32$
(Fig.~\ref{GroundState}),
the unit reciprocal lattice vectors for the FLL
are
\begin{eqnarray*}
	\vec{K}_1 = 2\pi [\frac{1}{8},-\frac{1}{8}],
	~~~
	\vec{K}_2 = 2\pi [0,\frac{1}{4}].
\end{eqnarray*}
In the FLL phase, $S(\vec{k}_\perp,k_z \keq 0)$ has $\delta$-function Bragg
peaks at $\vec{k}_\perp \keq \vec{K}(m,n) \keq m\vec{K}_1 \kp n\vec{K}_2$
(m,n=0,$\pm$1,$\pm$2,$\pm$3,..). The vectors $\vec{K}$ are located
within the first
Brillouin-zone. When the FLL melts, the Bragg peaks are smeared out. The
lowest temperature $T$ where $S(\vec{K}, k_z \keq 0)$ vanishes, thus
defines the FLL melting temperature $T_m$. For simplicity, we consider
only the structure function  $S(\vec{Q}) \!\equiv\! S(\vec{K}_2, k_z \keq 0)$.

\subsection{Flux-line cutting and intersection}
The issue of flux-line cutting  and flux-line and entanglement, as well as
the suggested possible resulting vortex states originating from the latter,
such as the analogs of $2D$ Bose superfluids and supersolids, and even
topological vortex glass states, have been issues of considerable controversy
over the last years. In particular, the effect of entanglement on the FLL
melting transition and the statistical mechanics of the FLL, has received
considerable attention. \\
In principle,
flux-line entanglement could be responsible for the drop in $\Upsilon_z$ we
observe at the temperature $T_z$. It is of interest to correlate the amount
of ``close vortex-line encounters'' with the anomalies we obtain in the
specific heat. This will allow us to at least tentatively decide whether or
not flux lines start to intersect or cut at any of the temperatures $T_m$,
$T_z$, or $T_{Bc2}$. \\
Intersection and cutting of
flux lines tend to act as efficient modes of {\it disentanglement} of flux
lines. It is unlikely that a flux-line liquid phase that suffers large
amounts of thermally induced collisions between flux-line segments, which
in turn strongly indicates that flux-line cutting takes place, can also
sustain heavily entangled vortex configurations. Hence, if we can show that
the amount of close vortex-segment encounters is substantial at the
temperature where $\Upsilon_z$ goes to zero, we may at least tentatively
conclude that the drop in $\Upsilon_z$ is {\it not} due to entanglement of
flux-lines.\\
To study the intersection and or cutting of  flux lines, we define a
parameter $\rho$, which is a measure of the amount of flux-line cutting and
intersection that takes place in the flux-line liquid.
\begin{eqnarray}
	\rho \equiv \frac{N_{int}}{N_{\phi}},
\label{rho}
\end{eqnarray}
where $N_{int}$ is the total number of unit cells having 4 or 6 vortex
segments penetrating their plaquette, and $N_{\phi}$ is the number of
field induced flux-lines in the system. Note that with this definition,
we only consider $\rho$ for the case of finite fields. This quantity
gives us some intuition on how well-defined we may expect individual
field-induced flux lines to be in the liquid phase. \\
In the Villain Model, each plaquette can carry
at most one vortex segment. This is because the vorticities are defined in
terms of gauge-invariant phase-differences, defined on the interval
$j_\mu(\vec{r}) \!\in\! [-\pi,\pi)$. The distribution of vortex segments
$\vec{n}(\vec{r})$ has no divergence
($\sum_{\mu=x,y,z} \nabla_\mu n_\mu (\vec{r}) = 0$).
Thus, a single unit cell can only carry 0, 2, 4, or 6 vortex elements.
If a unit has more than 2 plaquettes that are pierced by vortex segments,
then that corresponds to one of the three following situations:\\
(1) an intersection between two different flux lines, or \\
(2) an intersection between a flux line and a closed vortex loop, or \\
(3) an intersection between two different closed vortex loops.\\
Inside a unit cell, it is impossible to decide which possibility is
realized. In principle, $\rho$ as defined above counts all of these
possibilities, while flux-line intersection and flux-line cutting correspond
to case 1).

\section{Results, zero magnetic field}
We now present the results of our Monte-Carlo simulations, and consider
first the case of zero magnetic induction ${\bf B} = 0$.
{}From now on we measure the {\em specific heat per site $C$ in units of $k_B$,
the helicity moduli $\Upsilon_x$ in units of $J_0$, $\Upsilon_z$ in units of
$J_0/\Gamma^2$, and the temperature T in units of $J_0/k_B$}. \\
Two values of $\Gamma=\lambda_c/\lambda_a$, the anisotropy parameter, are
considered: $\Gamma =1$ and $\Gamma=3$. We will present results for the
quantities relevant to the zero-field case, namely the helicity modulus, the
vortex-loop distribution function, and the specific heat.  An important
point is that we need to, in the zero-field case, to be able to
correlate the temperature at which  the helicity modulus $\Upsilon_z$
vanishes, with the temperature at which the loop-distribution function $D(p)$
qualitatively changes  behavior from an exponential dependence on the
loop-perimeters to algebraic dependence on the loop-perimeters. Moreover,
both these features must be correlated with the temperature where an
anomaly in the specific heat is found, as discussed in Section II.D.
For arbitrary anisotropy $\Gamma$, the system should only have one
single phase-transition, the normal metal-superconductor transition.

\subsection{Helicity modulus}
\begin{figure}[htbp]
\psfig{figure=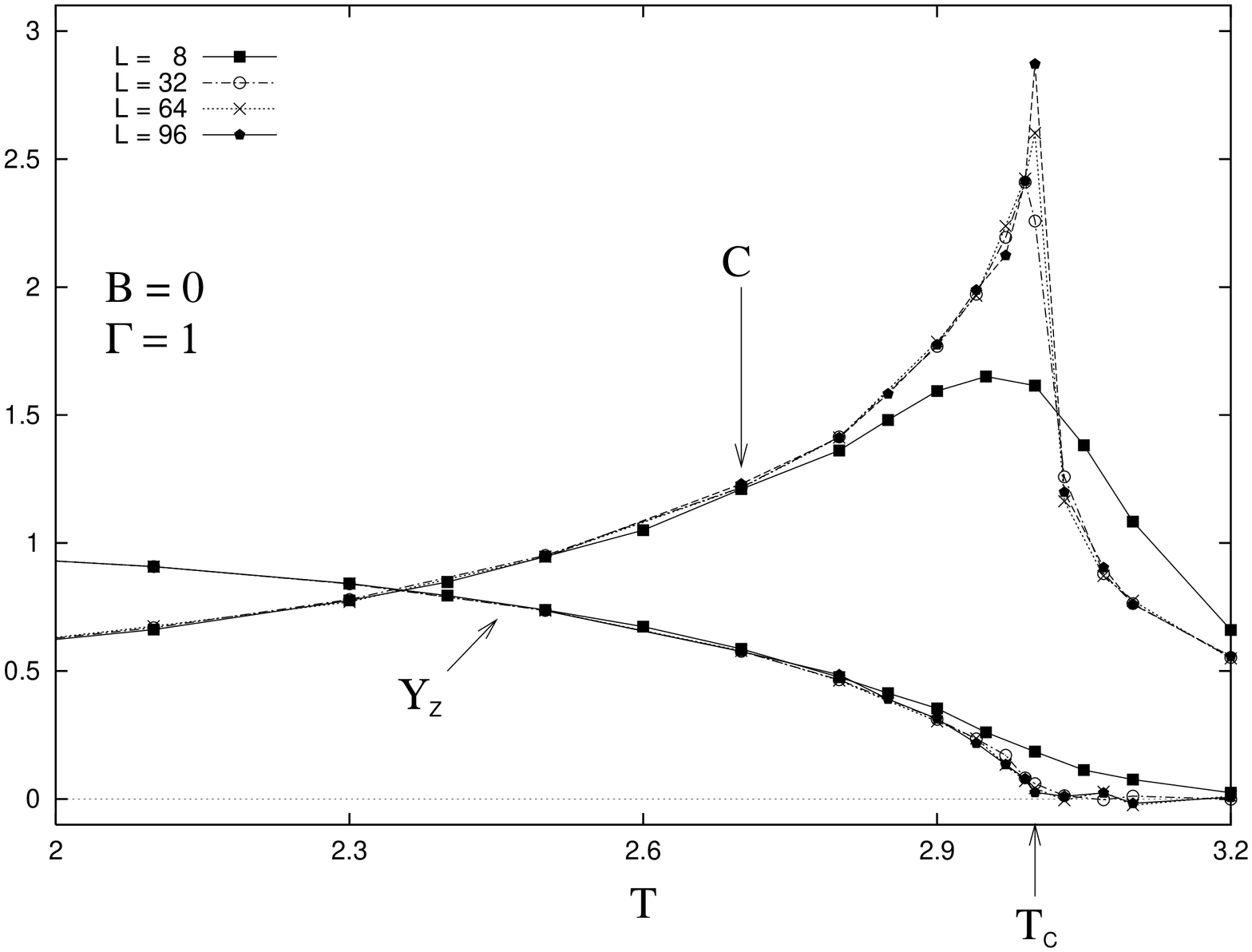,height=8cm,width=8cm,angle=270}
{\small FIG.~\ref{B0A1}. Helicity modulus $\Upsilon_z$ and the specific
	heat per site $C$ versus temperature, for $B \keq 0 $, $\Gamma$=1
	(isotropic) and system sizes L=8,32,64,96.}
\refstepcounter{figure}
\label{B0A1}
\end{figure}

The results for the helicity modulus $\Upsilon_{\mu}$, Eq. \ref{helmod},
in zero magnetic field and in the isotropic case $\Gamma=1$, are shown
in Fig. \ref{B0A1}.
We have confirmed that all moduli
$\Upsilon_x, \Upsilon_y, \Upsilon_z$, are equal in this case, and
therefore only exhibit $\Upsilon_z$. The results are  shown for the
system sizes $L=8,32,64,96$. Note that the drop in $\Upsilon_z$ becomes
sharper as the system size increases, and the value of $T$ where
$\Upsilon_z$ appears to vanish, becomes smaller. When $\lambda \to \infty$,
$\Upsilon_z$, which is the stiffness of the phase of the superconducting
order parameter to a twist, is proportional to the superfluid density
$\rho_s$. (For a finite $\lambda$ this identification no longer holds, as
emphasized in \cite{Teitel:L94}). We have found from the numerics that
$\Upsilon_z \sim | T-T_c | ^{2/3}$, consistent with the Josephson
scaling relation, $\rho_s \sim |T-T_c|^{\nu}$, where $\nu$ is the
correlation length exponent $\nu=2/d=2/3$.

\begin{figure}[htbp]
\psfig{figure=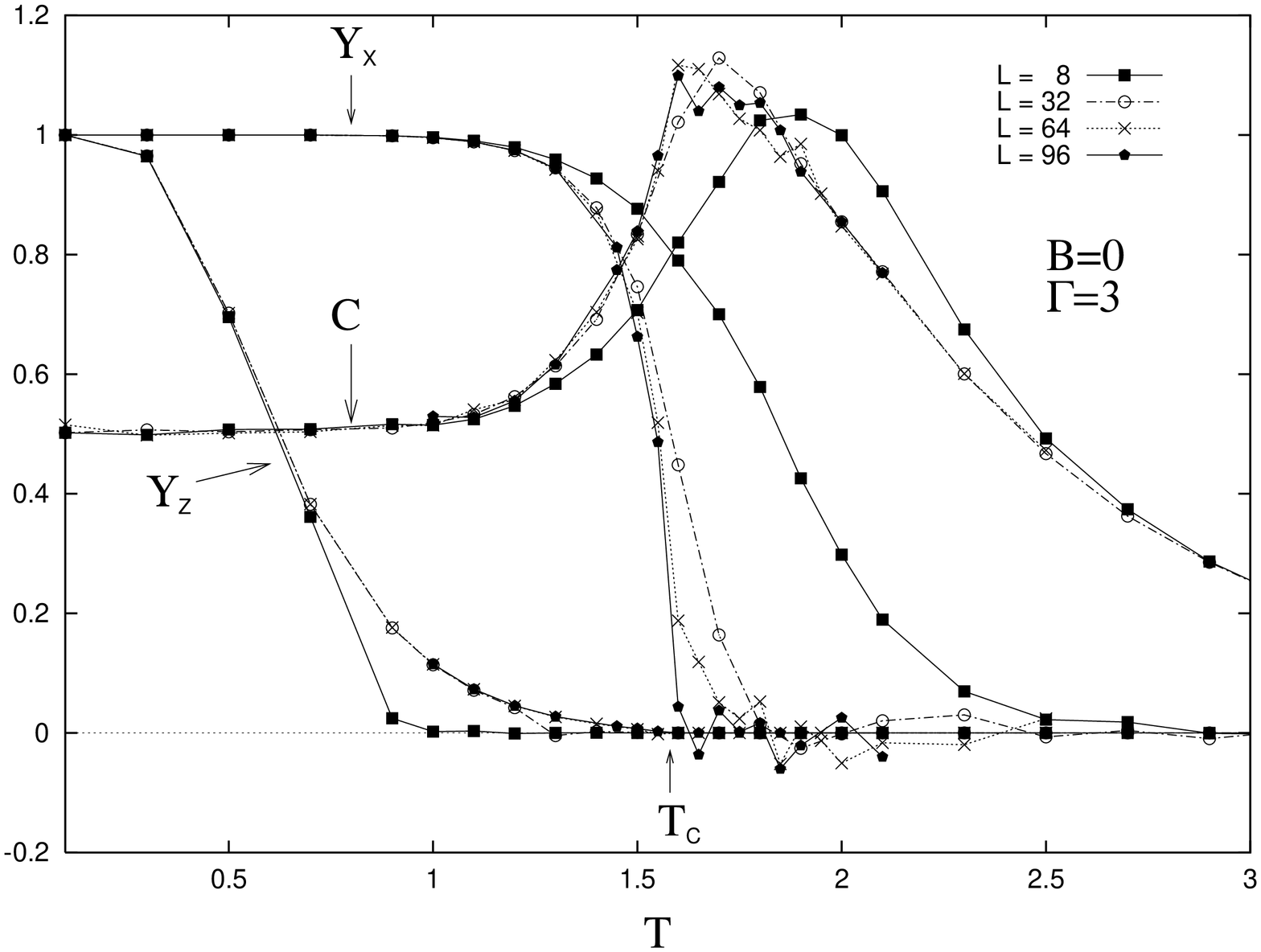,height=8cm,width=8cm,angle=270}
{\small FIG.~\ref{B0A3}. Helicity moduli $\Upsilon_x$, $\Upsilon_z$ and the
	specific heat per site $C$ versus temperature, for $B \keq 0 $,
	$\Gamma$=3 and system sizes L=8,32,64,96.
	For increasing system size $L$, $T_x$ approaches $T_c=1.57$ from
	above, and $T_z$ increases approaching $T_c$ from below. The layered
	system has only one phase transition.}
\refstepcounter{figure}
\label{B0A3}
\end{figure}

Due to the identification $\Upsilon_z \sim  \rho_s$, we conclude that
the vanishing of the helicity modulus corresponds to the normal
metal-superconductor transition.
The transition occurs at the temperature $T=3.0$ in units of $J_0/k_B$.  \\
The results for the helicity modulus in the anisotropic case $\Gamma=3$
are shown in Fig.\ref{B0A3}. The situation at first glance appears
considerably more complicated than in the isotropic case, despite our
expectations that the physics
basically should be the same as in the isotropic case, cf. our discussion in
Section II.D. A striking feature is that for $\Gamma=3$, the helicity modulus
$\Upsilon_z$ appears to vanish distinctly below the temperatures at which the
helicity moduli $\Upsilon_x,\Upsilon_y$ vanish. ($\Upsilon_x$ and
$\Upsilon_y$ turn out identical in all our simulations, and we therefore
only exhibit $\Upsilon_x$). Note however that there is an important
finite-size effect in the results: As $L$ increases, $\Upsilon_x$ vanishes
at progressively lower temperatures while $\Upsilon_z$ vanishes
at progressively higher temperatures. As $L$ increases, $\Upsilon_z$
and $\Upsilon_x$ appear to approach zero at the {\it same temperature}.
Due to the limitations in available system sizes, we have not been
able to perform the simulations at higher anisotropies
than $\Gamma=3$ {\it for the zero-field case}. At lower anisotropies
$1 < \Gamma < 3$, the same finite-size effect as described above is seen.
For lower values of the anisotropy it is also more obvious that the two
temperatures at which $\Upsilon_z$ and $\Upsilon_x$ vanish,
approach each other with increasing system size $L$. For the anisotropic
case $\Gamma=3$, the transition occurs at $T=1.57$ in units of $J_0/k_B$.

\subsection{Loop distribution}
\begin{figure}[htbp]
\psfig{figure=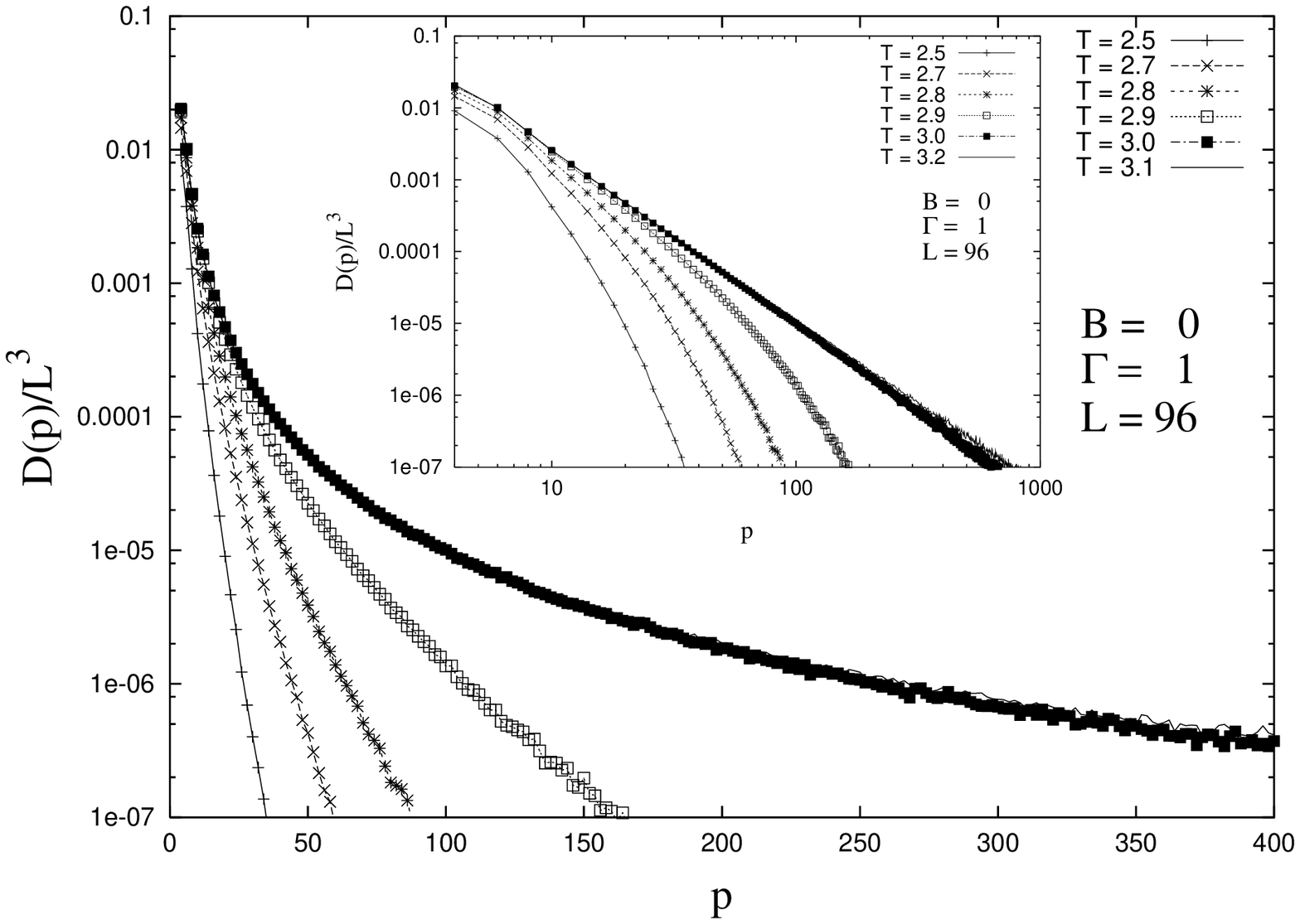,height=8cm,width=9cm}
{\small FIG.~\ref{B0A1.Dp}. The distribution function $D(p)/L^3$ of
	thermally excited, non-field induced, closed vortex loops per site
        as a function of perimeter p, for $B \keq 0 $, $\Gamma$=1, L=96 and
        various temperatures. It has been normalized this way to facilitate
	comparison between different system sizes. For $T < T_c \simeq 3.0$,
        $D(p)$ is best fit to an exponential decay. For $T=T_c$, $D(p)$ is
        fits an algebraic decay $D(p) \sim p^{-\alpha}$ with exponent
        $\alpha \sim 3$ excellently, indicating an Onsager loop-transition
        at $T=T_c$. The inset shows $D(p)$ on a log-log plot. The slope
        of the straight line, obtained at $T=T_c$, is $\sim -3$. At $T < T_c$
        the curves show a marked downward curvature, indicating a
        faster-than-algebraic decay of $D(p)$. This point is discussed in
        section II.F.}
\refstepcounter{figure}
\label{B0A1.Dp}
\end{figure}

In order investigate the excitations responsible for destroying the
superconducting phase-coherence and the superfluid stiffness as evidenced
by our results for $\Upsilon_z$, we probe the amount of
closed vortex-loops that are thermally excited in the superconductor model
at the temperature where $\Upsilon_z$ vanishes. We first discuss the
isotropic case $\Gamma=1$.\\
The results for $D(p)$ are shown in Fig. \ref{B0A1.Dp}, for the largest
system we have considered, $L=96$.
The figure shows $D(p)$ as a function of the loop perimeters $p$, for various
temperatures
in the range $T \in [2.3,3.3]$. Recall from above that in the isotropic
case, the helicity modulus vanished at $T \approx 3.0$.

{}From Fig. \ref{B0A1.Dp}, particularly from the inset of this figure,
we observe a qualitative change in $D(p)$ precisely at the temperature
$T=3.0$. The inset shows the distribution function on
a log-log plot, and it is seen that the decay is faster-than-algebraic
for $T<T_c$ while it is a precise power law with exponent $\alpha$ in
good agreement with a scaling analusis assuming that the vortex-loop blowout is
a critical point.  We attribute the slight deviation in the exponent
$\alpha$ between the simulations and the theory as due to the presence
of vortex loops of more complicated shapes, such that the circumference of
the loops do not all scale with their diameter.
We have tentatively suggested an exponential decay in the low-temperature
phase,
but this is not unambiguous. However, our main point is that for $T<T_c$
the decay cannot be a power law, while at $T=T_c$, the power law we find is
precisely the same as the one we predict analytically assuming that
the vortex-loop blowout is in fact a critical point.\\

\begin{figure}[htbp]
\psfig{figure=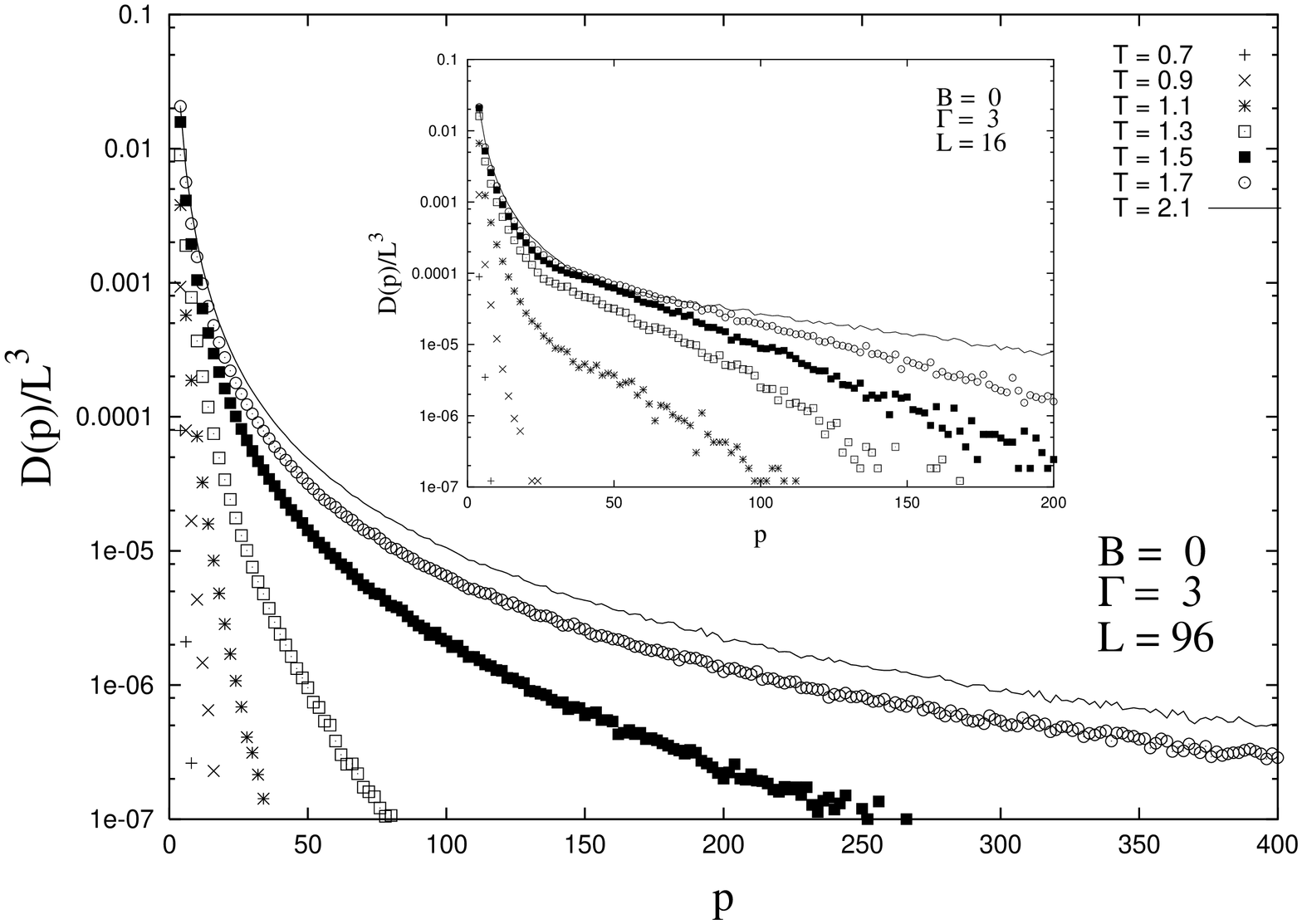,height=8cm,width=8cm,angle=270}
{\small FIG.~\ref{B0A3.Dp}. The distribution function $D(p)/L^3$ of
	thermally excited closed vortex loops per site as a function of
	perimeter $p$, for $B \keq 0 $, $\Gamma$=3, $L=96$ and various
	temperatures. For $T < T_c \simeq 1.6$, $D(p)$ is best fit to
	an exponential decay. For $T = T_c$, $D(p)$ is best fit to an
	algebraic decay, $D(p) \sim p^{-\alpha}$, with $\alpha \sim 3$
	consistent
	with critical scaling of section II.F. The inset shows the same
	figure for L=16. Note that for L=16, $D(p)$ incorrectly shows
	algebraic decay for $T = 1.3 < T_c$. This finite-size effect is
	discussed in Section II.D.}
\refstepcounter{figure}
\label{B0A3.Dp}
\end{figure}

Hence, based on the above, we conclude that at $T=3.0$, a sharp
phase-transition
occurs from a low-temperature phase where closed vortex-loops are confined
to some typical size $L_0(T)$, to a high-temperature phase where closed
vortex loops of all sizes up to and including the system size, exists.
Thus, in the isotropic case we have been able to precisely correlate
the drop in $\Upsilon_z$
with a vortex-loop blowout, and from the previous paragraph we must
identify this as the superconductor-normal metal transition.\\
The loop-distribution function for the anisotropic case $\Gamma=3$ is shown
in Fig. \ref{B0A3.Dp}.
Due to the drop in the critical temperature of the
system, we now show $D(p)$ as a function of $p$ for temperatures in the
range $T \in [0.7,2.1]$. Again, we observe a qualitative change in the
behavior of $D(p)$ from exponential decay to algebraic decay, at a
temperature $T \approx 1.6$, which correlates almost perfectly with
the temperature $T=1.57$ at which the helicity moduli $\Upsilon_z$ and
$\Upsilon_x$ vanish for $\Gamma=3$. If we fit $D(p) \sim p^{-\alpha}$ at this
temperature, we again find the exponent $\alpha=3$, as in the isotropic case.\\
So far, our expectations based on the discussion in Section II.D are
borne out. To illustrate the point further, in the inset of
Fig. \ref{B0A3.Dp}, we show the distribution function $D(p)$ for
$\Gamma=3$ and the same range of temperatures, for the smaller
system $L=16$. The important difference between these two cases is
that for $L=16$, algebraic decay of $D(p)$ appears to persist down to
lower temperatures than for $L=96$. For larger systems the vortex-loop
blowout is suppressed due to the fact that the interplane coupling is
allowed to renormalize further without being cut off by a small system
size. Hence, what appeared to be a separate vortex-loop blowout at a
low temperature $T=1.1$ for $L=16$, has been pushed up to the correct
temperature $T \approx 1.6$ in the larger system $L=96$, as discussed in
Section II.D.

\subsection{Specific heat}
We now present  our results for the specific heat in zero magnetic field
for the system sizes $L=8,32,64,96$, and consider first the isotropic
case $\Gamma=1$. \\
The specific heat in the isotropic case is shown in Fig. \ref{B0A1}.
The anomaly in the specific heat clearly correlates with the temperature
where the superfluid stiffness vanishes, which in turn correlates
precisely with the temperature where the vortex-loop blowout is observed.
As the system size increases the anomaly clearly also becomes sharper.
We have shown that the peak in the specific varies in very good agreement
with $\ln(L)$ as the system size is increased, further indicating a genuine
thermodynamic phase transition. The shape of the specific heat curve
has a typical $XY$-behavior for this extreme type-II superconductor
$(\lambda \to \infty$). This agrees with previous results found
by Dasgupta and Halperin  for the lattice superconductor model
\cite{Dasgupta:L81}. (Note that this contrasts sharply with the specific
heat results indicating an {\it inverted} XY-transition found by the same
authors for finite, small $\lambda$ in the isotropic case). Qualitatively,
our results also agree well with the specific heat measurements on YBCO of
Schilling {\em et al.} \cite{Schilling:B95}. \\
The specific heat for the anisotropic case $\Gamma=3$ is shown in
Fig.~\ref{B0A3} for the system sizes $L =8,32,64,96$.
The situation again at first glance
appears to be more complicated than the isotropic case. However, as in the
isotropic case, the peak in the specific heat anomaly increases with system
size. Moreover, the temperature of the peak in the specific heat is reduced
as the system size is increased, approaching the temperature $T = 1.57$ at
which both $\Upsilon_z$ and $\Upsilon_x$ vanish, and where the vortex-loop
blowout appears to take place. \\
Hence, our simulations of the above three quantities $\Upsilon_z$, $D(p)$,
and specific heat correctly capture the physics
that even in the very anisotropic case, only
one single phase transition occurs in the lattice superconductor model
in zero magnetic field.
The fact that our simulations do not reveal an artificial zero-field
decoupling transition due to finite-size effects, makes us confident that
our simulations are able to capture the subtle zero-field physics
correctly in the anisotropic case. This is a necessary prerequisite for
being able to extract meaningful results from our finite field-simulations,
to which we now turn.

\section{Results, finite magnetic field}
Next, we present results for finite magnetic induction. We will consider
the filling fraction $f=1/32$ corresponding to $32$ ab-plane plaquettes
per field-induced flux line in the ground state, depicted in
Fig. \ref{GroundState}. Again, we are primarily
interested in correlating temperatures where the helicity moduli vanish
with temperatures where anomalies in specific heat occur, as well as with
temperatures where we see a qualitative change in the distribution of
non-field induced closed vortex-loops. It is
also of interest to correlate these phenomena with the melting of the FLL
as evidenced by a drop in the structure function at low-order Bragg-peaks.
Furthermore, due to the presence of field-induced flux lines, it is also
of interest to monitor the amount of flux-line cutting occuring in the
system. This has bearing on the amount of entanglement of flux lines the
molten vortex phase can sustain. For finite-fields we present results for
helicity moduli, vortex-loop distribution, specific heat, structure
function and flux-line cutting for the two values of the anisotropy
parameter, $\Gamma=1$ and $\Gamma=3$.

\subsection{Structure function}
\begin{figure}[htbp]
\psfig{figure=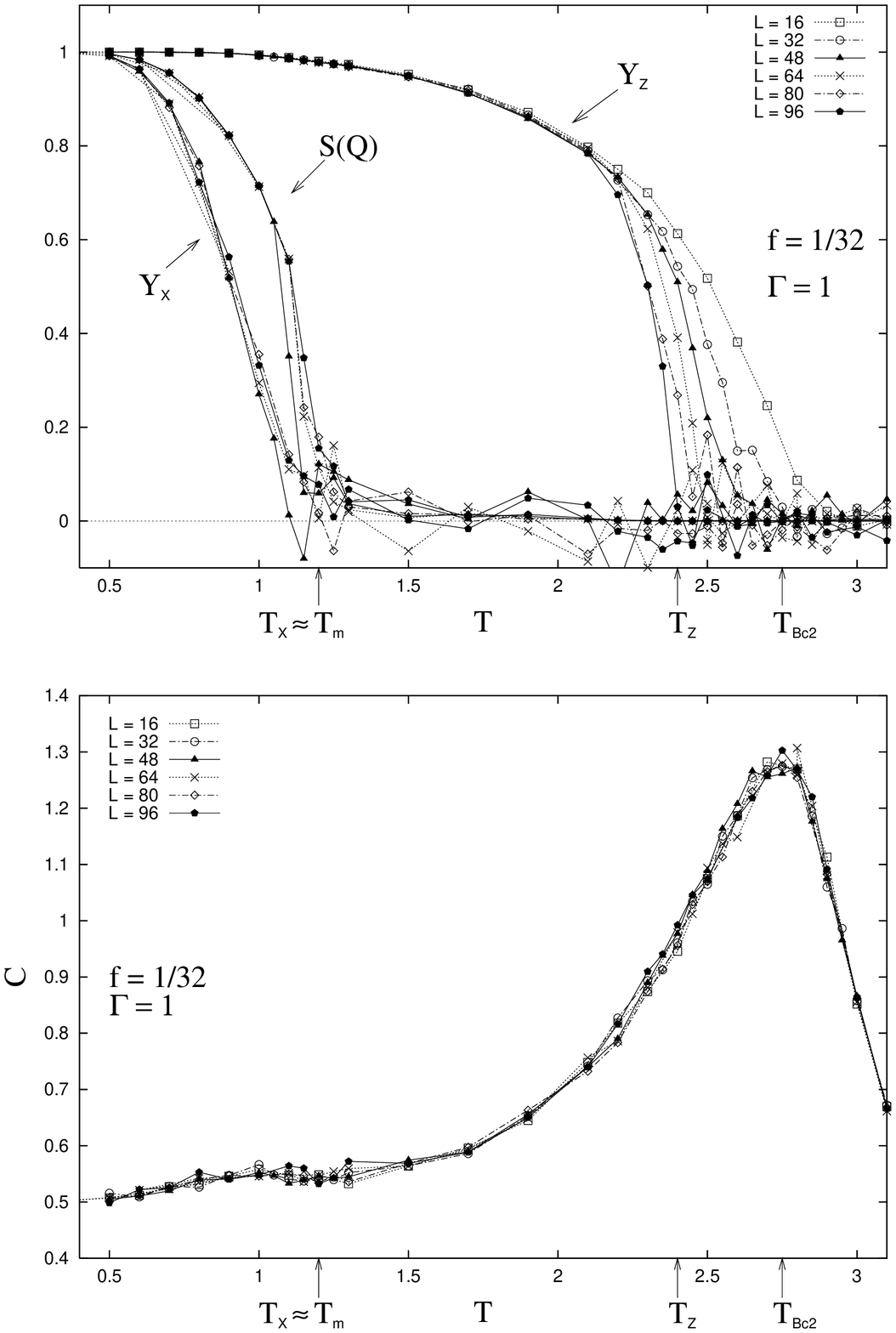,height=12cm,width=8cm,angle=180}
{\small FIG.~\ref{B32A1.GGS.Size}. Top panel: Helicity moduli $Y_x$ and $Y_z$
	and  structure factor $S(\vec{Q})$ versus temperature, for
	$f=1/32$, $\Gamma$=1 and $L=16,32,48,64,96$. Note how for
	increasing system size, $T_m$ increases slightly, and $T_z$
	decreases markedly. Lower panel: The specific heat per site
	$C$ versus temperature for $f=1/32$, $\Gamma$=1 and
	$L=16,32,48,64,96$. }
\refstepcounter{figure}
\label{B32A1.GGS.Size}
\end{figure}

The results for the structure function $S({\vec {Q}})$ are shown in the
top panel of Fig. \ref{B32A1.GGS.Size} for the isotropic case $\Gamma=1$,
for the reciprocal lattice vector $\vec{Q}=(\vec{K}_2,k_z=0)$,
${\vec K}_2= 2 \pi[0,1/4]$, and for the system sizes $L=48,64,80,96$.
In the top panel of  Fig. \ref{B32A3.GGS.Size}, results for $\Gamma=3$
and for various system sizes $L=32,48$ are also shown.
The structure function exhibits a sharp drop at $T=T_m$, well below
the temperature where the phase-coherence in the direction of
the magnetic field vanishes. (We discuss the helicity modulus $\Upsilon_z$
in the next section.) Note that the structure function vanishes essentially
at the same temperature as the temperature where the helicity modulus
$\Upsilon_x$ vanishes.

What this indicates is that the filling fraction we have used in our
simulations, $f=1/32$, is not sufficiently small to study the melting of
the FLL in the continuum limit. The FLL has not thermally `depinned' from
the numerical mesh at the melting transition. Therefore, our estimate for
the FLL melting temperature as obtained in these simulations is too high.
However, from our earlier work on the moderately anisotropic
Lattice London model \cite{Nguyen:L96}, we know that $f=1/48$ suffices to
produce a thermal `depinning' temperature of the FLL off of the
numerical mesh at a temperature distinctly below the observed melting
temperature of the FLL. Therefore, we expect that the present estimate for
$T_m$ should be quite good; only a minor reduction of the filling fraction
below $f=1/32$ is expected to suppress the `depinning' temperature $T_x$
below the melting temperature $T_m$. It is also possible that the
commensuration effect might tend to overestimate a first order character
of the FLL melting transition, should such a result be found. \\
The reduction in the structure function below the melting temperature is
due to the Debye-Waller factor
\begin{eqnarray}
	S({\vec Q},T) = S({\vec Q},0) ~~ \exp(- G^2 <u^2>)
\label{dwfac1}
\end{eqnarray}
For a triangular lattice, we find right below the melting transition
\begin{eqnarray}
	S({\vec Q},T) = S({\vec Q},0) ~ \exp(- \frac{8 \pi}{3} c_L^2)
\label{dwfac2}
\end{eqnarray}
where $c_L$ is the Lindemann-ratio. In our simulation, we find that the
DW-factor is $0.6$ right below the melting transition, and hence
$c_L=0.24$. Essentially the same result is found for the isotropic case,
and is in reasonable agreement with the value $c_L=0.4$ used in the best
calculation so far to estimate the position of the FLL-melting line in $BSCCO$
and $YBCO$ by employing the simple Lindemann-criterion in conjunction with
the highly nontrivial fluctuation propagator found from anisotropic and
non-local elastic theory of the flux-line lattice \cite{HPS:B39}.

\subsection{Helicity modulus}
\begin{figure}[htbp]
\psfig{figure=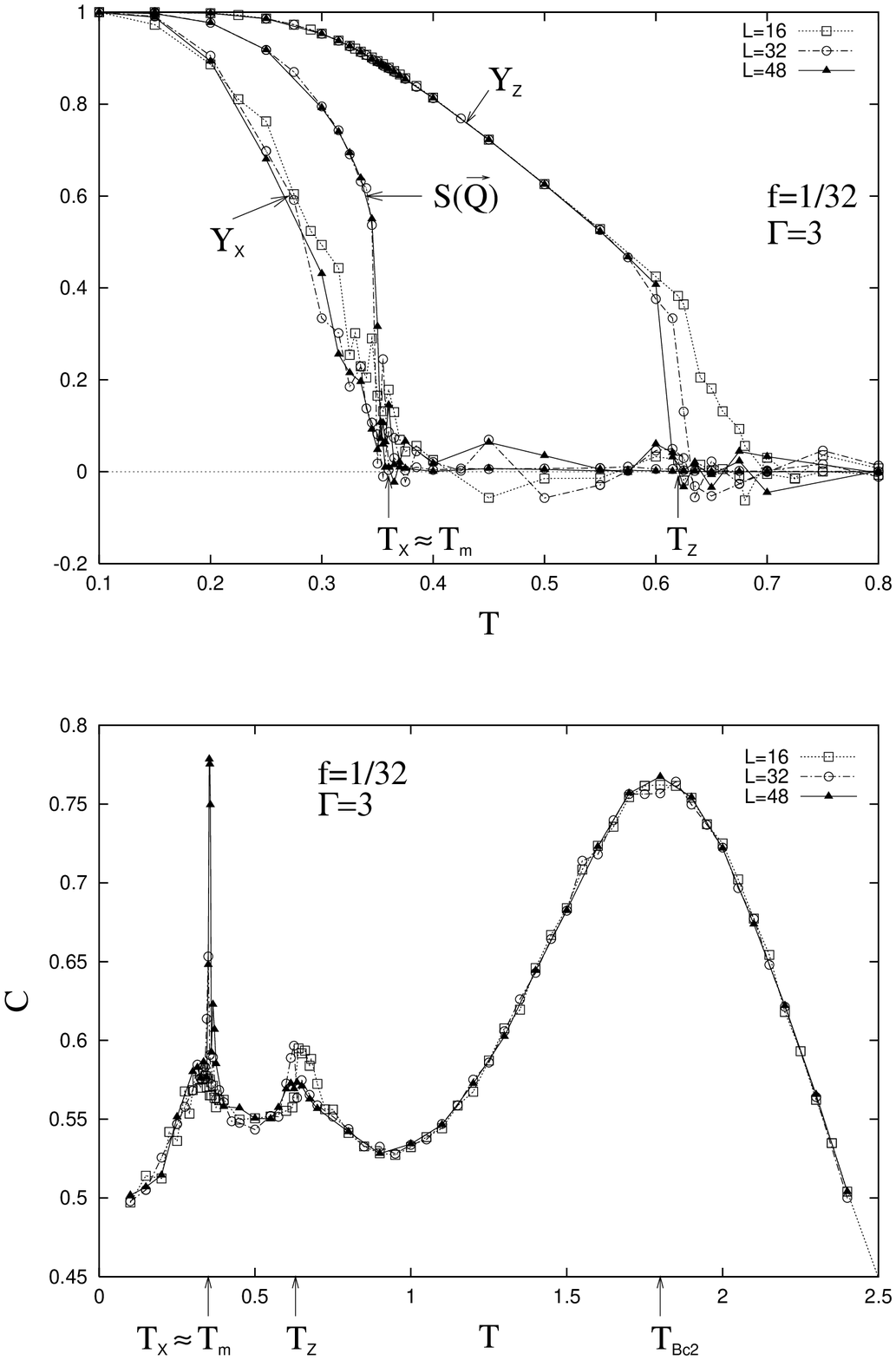,height=12cm,width=8cm,angle=180}
{\small FIG.~\ref{B32A3.GGS.Size}. Top panel: Helicity moduli
	$Y_x$ and $Y_z$ and  structure factor $S(\vec{Q})$ versus
	temperature, for $f=1/32$, $\Gamma$=3 and $L=16,32,48$.
	Note how for increasing system size, $T_m$ increases slightly
	and $T_z$ decreases markedly.
	Lower panel: The specific heat per site $C$ versus temperature
	for $f=1/32$, $\Gamma$=3 and $L=16,32,48$. Note the two separate
	specific heat per site anomalies at $T_x=T_m$ and $T_z$. The
	specific heat peak at $T_m$ becomes more prominent with increasing
	system size, while the peak at $T_z$ actually decreases with
	increasing system size.}
\refstepcounter{figure}
\label{B32A3.GGS.Size}
\end{figure}

\begin{figure}[htbp]
\psfig{figure=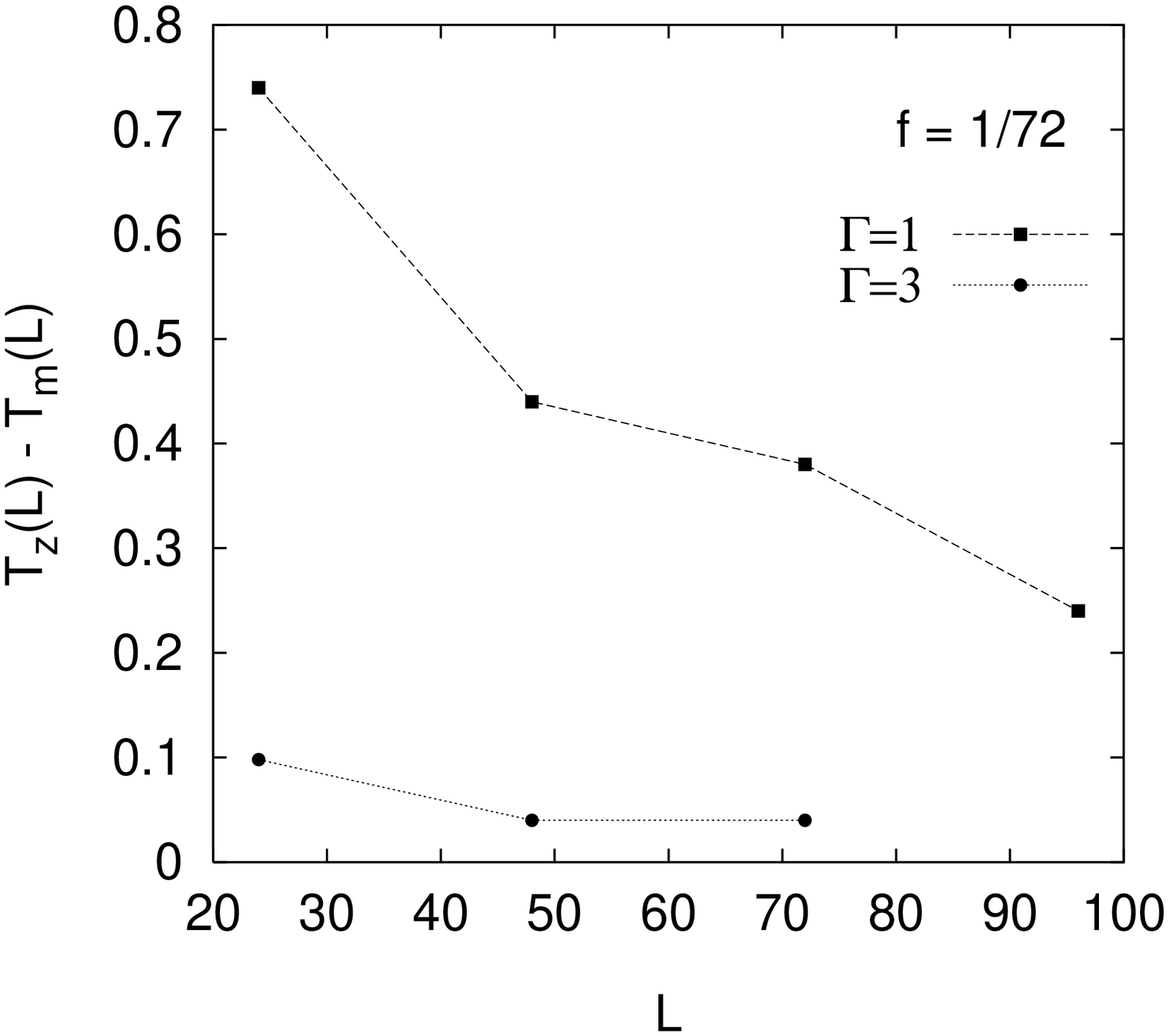,height=7.5cm,angle=270}
{\small FIG.~\ref{F72A1A3}.~The difference $T_z(L) - T_m(L)$ as a function
of L for $f=1/72$, $\Gamma = 1$ and $\Gamma = 3$. Note that this quantity
shows a monotonically (sublinear) decrease with $L$. Note also that the
relative decrease is larger for $\Gamma=1$ than for $\Gamma=3$. We attribute
this to a slower relaxation in the anistropic case than in the isotropic,
since the both $T_m$ and $T_z$ are smaller for $\Gamma=3$ than for $\Gamma=1$.}
\refstepcounter{figure}
\label{F72A1A3}
\end{figure}

The results for the helicity moduli $\Upsilon_z$ and $\Upsilon_x$
are shown for $\Gamma=1$ and system sizes $L=16,32,48,64,80,96$
in the top panel of Fig. \ref{B32A1.GGS.Size}. Likewise, similar
results for $\Gamma=3$ are shown in the top panel of Fig.~\ref{B32A3.GGS.Size}
for system sizes $L=16,32,48$. Note how the temperature $T_z$
appears to decrease monotonically with system size. \\
An important issue is how $T_z$ will continue to vary when the
system size is increased indefinitely. Since there is no obvious sign of
saturation in $T_z$ as $L$ increases, it could conceivably continue
to decrease in the liquid phase, until it reaches $T_m$.
Does this in fact happen, or is $T_z > T_M$ in the thermodynamic and
continuum limit? \\
To answer this question, we have performed simulations on systems
with filling fractions $f=1/72$, and computed $T_m$ and $T_z$
as a function of $L$. The point about going to lower filling fractions
is that we are approximating the continuum limit better.
It is becoming increasingly clear from numerical simulations
that when the $xy$-plane is discretized in order to do the
simulations one is introducing a long time scale into the problem:
There is a gap for moving vorticities from one unit cell to another.
When the filling fraction is lowered, the continuum approximation
is better approximated, and the relaxation time introduced by
discretization is lowered. Hence, for $f=1/72$ we are better able
to equilibrate the system. (Incidentally, we believe the reason
that no finite-size effect was seen in $\Upsilon_z$ in
Ref. \cite{Teitel:B93} was precisely that the simulations were not run
for a long enough time). The results of our simulations for this
cases is shown in Figure \ref{F72A1A3}, where we show $T_z(L)-T_m(L)$
for $\Gamma=1$ and $\Gamma=3$, for the case $f=1/72$.
It is seen thatv $T_z(L)-T_m(L)$ decreases monotonically as a function
of $L$.
Is it possible that $T_z$ could drop arbitrarily far below
$T_m$ when the anisotropy $\Gamma$ is increased indefinitely?
We believe that the answer for $\lambda = \infty$ is no, for the following
reason. \\

Recall our discussion in Section II.D, where it was shown that
in zero field, no vortex-loop blowout can take place below the
zero-field transition temperature, due to the renormalization of the
interplane coupling. In finite fields, this is quite different, since
the superconducting coherence length is limited by the {\it magnetic
length} as soon as the flux-lines start to fluctuate appreciably, cutting
off the renormalization of the interplane phase-coupling. Hence, we see
that $T_z$ drops well below $T_{Bc2}$, in finite fields, provided the
system is in the liquid-phase. In the flux-line {\it lattice} phase,
the coherence length is no longer limited by the magnetic length,
there is now phase-coherence throughout the sample (but of a more
complicated form then in the zero-field case). Hence, the
renormalization of the interplane coupling described in Section II.D
again becomes active, suppressing the vortex-loop blowout.
The flux-line lattice phase is therefore in some sense equivalent to
the zero-field case with regards to
a vortex-loop blowout, and a loss of phase-coherence along the
field direction cannot take place within the lattice phase. {\it Hence,
$T_z$ cannot drop far below $T_m$}, at least not for the case
$\lambda = \infty$.

In our previous work \cite{Nguyen:L96} we emphasized the importance of closed
loops, but suggested that the drop in $\Upsilon_z$ within the liquid phase was
a genuine phase-transition from a coherent vortex liquid characterized by
$\Upsilon_z \neq 0$, to an incoherent vortex-liquid characterized by
$\Upsilon_z = 0$. We strongly believe our present simulations on much larger
systems
show that this in fact is not the case: $T_z \to T_m$ as the system size
increases. However, the final word on the issue, particularly
for the isotropic case $\Gamma=1$, remains to be said. \\
Note also that, compared to the zero-field case $f=0$, $\Upsilon_z$ vanishes
at a considerably lower temperature when $f=1/32$. The helicity modulus
$\Upsilon_x$ vanishes at a finite temperature below $\Upsilon_z$.\\
The fact that $\Upsilon_x$ vanishes at a finite temperature is an artifact
of our discretizing the ab-plane. In the continuum limit, this helicity
modulus would be zero for any finite temperature when no physical pinning
of the flux lines is present. On the other hand, $\Upsilon_z$ has little
or no commensuration effects in it. \\
The above result indicates that phase-coherence across the sample along the
direction of the flux
lines is lost at a finite temperature $T_z$. Contrary to the zero-field
case, there are several possible explanations for this loss of phase
coherence when a finite field is present. One possibility is that a
vortex-loop blowout causes the loss of phase-coherence at $T=T_z$.
Another explanation  could be
that since the melting temperature $T_m$ is smaller than $T_z$, in FLL
liquid phase flux lines become entangled  thereby destroying the phase
coherence in the superconducting order-parameter along the direction of
the flux lines. A third explanation could be that transverse flux-line
meanderings and flux-line cutting causes loss of phase-coherence without a
resulting entanglement of flux-lines. We discuss these possibilities in turn.

\subsection{Loop distribution}
\begin{figure}[htbp]
\psfig{figure=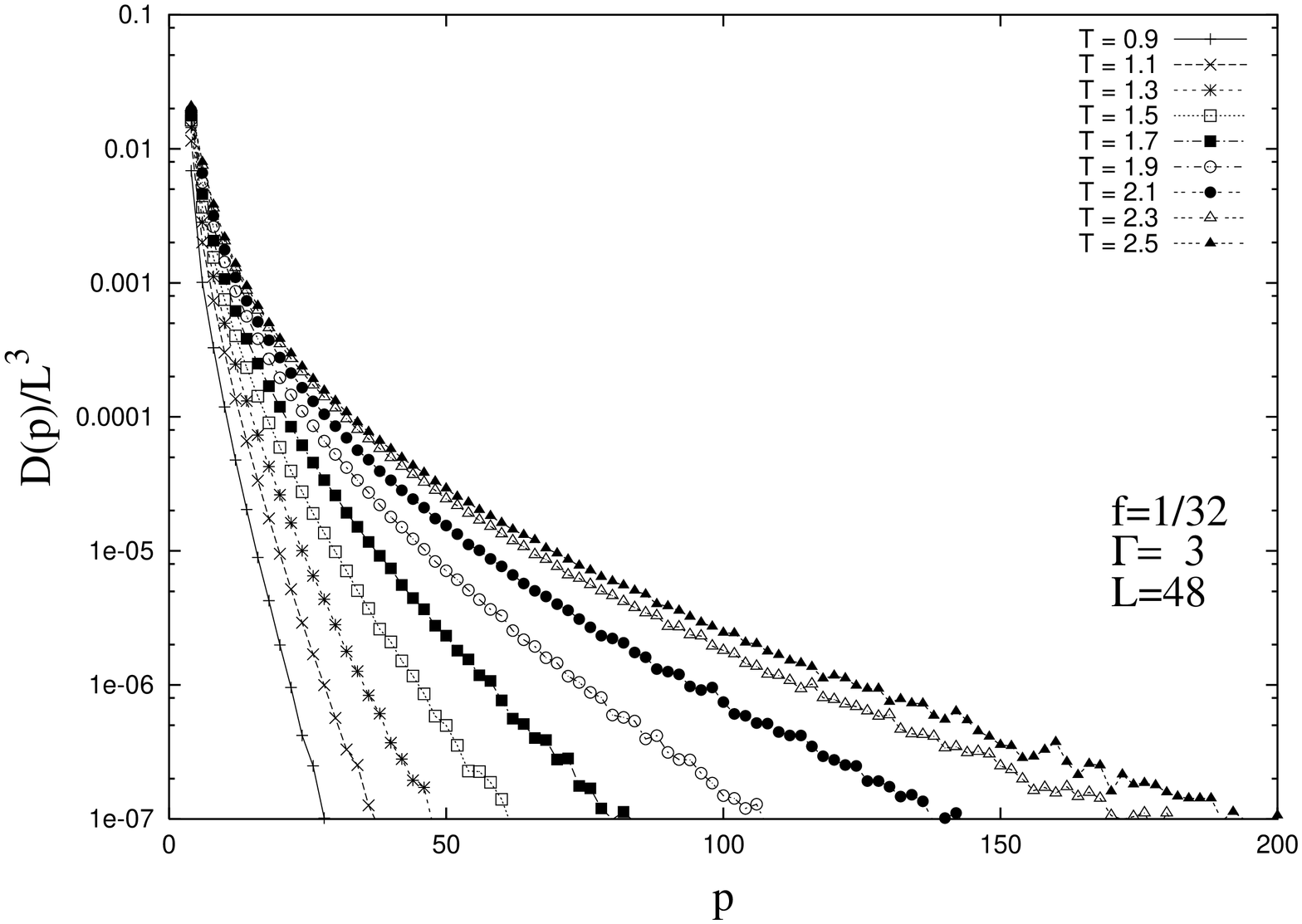,height=8cm,width=8cm,angle=270}
{\small FIG.~\ref{B32A3.Dp}. The distribution function $D(p)/L^3$ of
	thermally excited closed vortex loops as a function of perimeter
	$p$, for $f=1/32$, $\Gamma$=3, L=48 and various temperatures.
	For $T < T_{Bc2} \simeq 1.8$, $D(p)$ is best fit to an
	exponential decay. For $T>T_{Bc2}$, $D(p)$ is best fit to an
	algebraic decay, indicating a vortex loop blow out at
	$T_{Bc2} \gg T_z > T_m$.}
\refstepcounter{figure}
\label{B32A3.Dp}
\end{figure}

The results for the distribution of closed vortex-loops are shown
in Fig. \ref{B32A3.Dp}, in the temperature range $T \in [0.9,2.5]$,
for the anisotropy $\Gamma=3$ and the system size $L=48$.
This temperature range encompasses the melting of the FLL and the destruction
of phase-coherence along the direction of the magnetic field. A
feature which distinguishes the finite-field results for $D(p)$ from the
zero-field case, is that throughout the temperature range where the FLL
melts, $T_m\approx 0.38$ and where phase-coherence is lost,
$T_z \approx 0.65$, the distribution function $D(p)$ decays more rapidly
than at the critical point in zero field.
In the range $T < 0.7$ there is no obvious sign in $D(p)$ of  a cross-over to
algebraic decay as a function of loop-perimeters. We conclude that the
vanishing of $\Upsilon_z$ is not, in this case, associated with a
vortex-loop blowout on all length scales of the system. In other word, the
vanishing of $\Upsilon_z$ is not due to a finite-field counterpart of the
Onsager-loop blowout we found in zero field. We find a change in the behavior
in $D(p)$ from exponential decay to algebraic decay at a much larger
temperature $T \approx 2.8$ for $\Gamma=1$ and $T \approx 1.9$ for
$\Gamma=3$. The temperature range over which $D(p)$ changes behavior
is also considerably broader than in the zero-field case, indicating that
the vortex-loop blowout transition which was found to be sharp at zero field,
is replaced by  a crossover. \\
However, the interaction between closed vortex-loops and the flux-line
lattice  may be studied by considering the number of closed vortex-loops
with a diameter given by the magnetic length in the problem. This number
scales with $L_z$ at $T_z$, and thus in the thermodynamic limit there
are infinitely many such vortex-loops per flux-line at the temperature
where the FLL melts.

\subsection{Specific heat}
\begin{figure}[htbp]
\psfig{figure=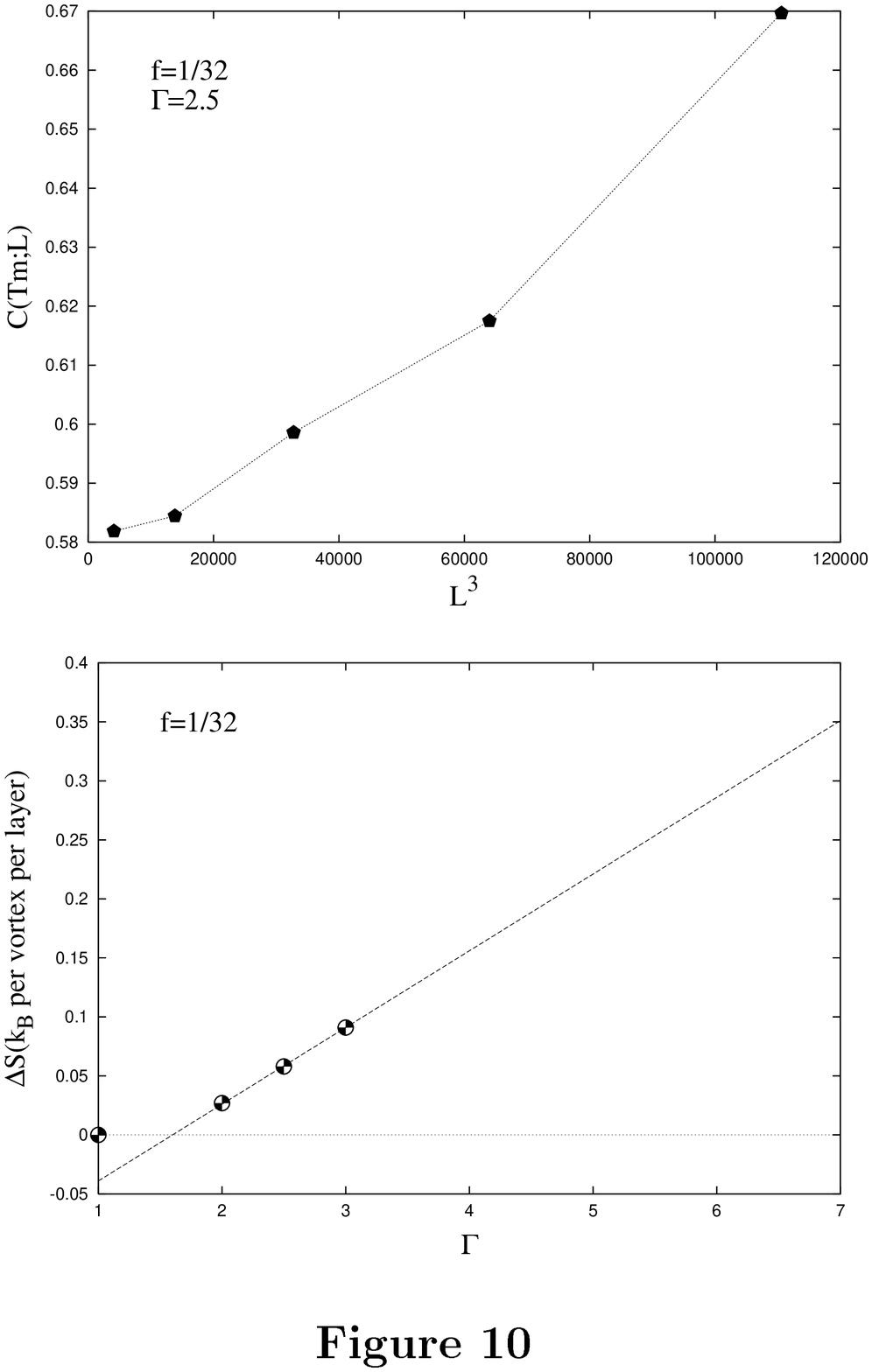,height=14cm,width=8cm,angle=0,clip=true}
{\small FIG.~\ref{Spec.Heat}. Top panel: The specific heat-maximum at the
	flux-line lattice melting transition, as a function of system size,
	for a given mass-anisotropy $\Gamma=2.5$. Such plots may be used
	to extract the entropy discontinuity at the melting transition,
	as explained in the text.  Bottom panel: Entropy discontinuity at
	the melting transition as a function of the mass-anisotropy. The
	first order character of the flux-line lattice melting is seen to
	increase rapidly with increasing $\Gamma$.}
\refstepcounter{figure}
\label{Spec.Heat}
\end{figure}

Our results for the specific heat are shown in the bottom panel of
Fig.\ref{B32A1.GGS.Size} for the parameters $f=1/32$, $\Gamma=1$
and $L=16,32,48,64,80,96$. Similar results for the anisotropic case
$\Gamma=3$ are shown in the bottom panel of Fig. \ref{B32A3.GGS.Size}.
It is clear that that the
lowest anomaly in the specific heat correlates with the melting of
the FLL.
There is also a broad
feature at a much higher temperature, $T_{Bc2} \approx 1.9$, which is roughly
equal to the temperature at which we see a sharp peak in the specific heat
in the zero-field case. \\
The feature at $T_{Bc2}$ is the remnant of the zero-field anomaly on the
specific heat, previously shown in Fig.\ref{B0A3}. This broad peak in the
specific heat is associated with the upper critical field $H_{c2}(T)$, and
our results show that in the extreme type-II
Villain-model, $H_{c2}(T)$ line is very steep close to the zero-field
$T_c$. As discussed in the previous section, $T_{Bc2}$ is also close to
the temperature at where the distribution function $D(p)$ of closed
vortex-loops changes behavior for exponential to algebraic decay.
Hence, we conclude that in the field-regime  corresponding to
$f=1/32$, the loop-transition appears close to the mean-field
$H_{c2}(T)$-line. \\
We caution the reader that $f=1/32$ is not a particularly low magnetic
field. If we estimate it for YBCO using the method of
Ref. \cite{Hetzel:L92}, it corresponds to  a magnetic field of the order
of $1T$. As recently emphasized by Te{\v s}anovi{\'c}
\cite{Tesanovic:B95} and Nguyen {\it et al.} \cite{Nguyen:L96} such
magnetic fields may not be relevant for discussing  the low-field
experiments of Zeldov {\it et al.} \cite{Zeldov:N95}. Therefore, our
simulations do not address the issue of the fate of the zero-field
loop-transition in asymptotically low magnetic fields, of the order of
$100G$ and below. Neither do the simulations address how this finite-field
counterpart of the zero-field vortex-loop transition interacts with the
melting transition of the FLL and with the loss of phase-coherence along
the field direction, in this low-field regime relevant for discussing the
results in Ref. \cite{Zeldov:N95}. \\
We next turn to a discussion of the anomalies at the two lower temperatures
$T_m$ and $T_z$, and base our discussion on the lower panel of
Fig. \ref{B32A3.GGS.Size}.
It shows the specific heat for the filling  $f=1/32$
and anisotropy $\Gamma=3$, for various system sizes $L=16,32,48$.
The feature in the specific heat at $T=T_m$ is clearly associated with
FLL melting. One notable feature in the specific heat anomaly at $T=T_m$
is that its peak scales as $L^3$, characteristic of a {\it first  order}
melting transition \cite{Dasgupta:L81}. This follows from the fact that a
first order phase-transition is generally characterized by
coexistence of two phases at the transition: One low-energy
ordered phase and one high-energy disordered phase. Thus there is a
discontinuity in the internal energy of the system at the transition
temperature, {\it and a delta-function peak} in the specific heat. On a
finite system the delta-function peak is converted to a peak of
order $L^d$, where $L$ is the linear dimension of the system, and
$d$ is its dimensionality. In fact, the coefficient of the $L^d$
term is the discontinuity in the entropy of the system, at the
transition  \cite{Dasgupta:L81}, so for the $3D$ case we obtain
\begin{eqnarray*}
   C = const +
	\frac{L^3}{4} \left ( \frac{\Delta S}{ k_B L^3} \right )^2.
\end{eqnarray*}
Thus we may use finite-size scaling of the specific heat to extract
$\Delta S$, or equivalently, the latent heat of the melting transition.
We find
\begin{eqnarray*}
   \Delta S & \approx & 0.00 ~ k_B/{\rm{vortex ~ per ~ layer}},~~~~\Gamma=1.0
\nonumber \\
   \Delta S & \approx & 0.03 ~ k_B/{\rm{vortex ~ per ~ layer}},~~~~\Gamma=2.0
\nonumber \\
   \Delta S & \approx & 0.05 ~ k_B/{\rm{vortex ~ per ~ layer}},~~~~\Gamma=2.5
\nonumber \\
   \Delta S & \approx & 0.10 ~ k_B/{\rm{vortex ~ per ~ layer}},~~~~\Gamma=3.0
\end{eqnarray*}
The results for a finite-size scaling of the specific heat per site
$C$ and the
entropy discontinuity at the melting transition, are shown in
Fig. \ref{Spec.Heat}. $\Delta S$ is seen to increase rapidly with $\Gamma$.
This is expected on general grounds, since the flux-line liquid in
a very anistropic superconductor is expected to exhibit more disorder
than in an isotropic case due to the more flexible nature of
{\it individual} flux lines in layered compounds. In our opinion, a reduction
of $\Delta S$ with increasing $\Gamma$ would be unphysical, at least in
situations where the superconductor can be viewed as an anisotropic continuum.
The $2D$ case is in some sense a singular limit, as discussed in Section II.D.

Although it is expected on general grounds that the melting of the FLL is
first order also in the isotropic case, it appears to be too small to detect
in our simulations. However, by extrapolation we may easily extract entropy
jumps of the correct order of magnitude seen in calorimetric data on
$YBCO$ with $\Gamma \approx 8$ \cite{Schilling:N96}, since $\Delta S$
is expected to grow with further increase in $\Gamma$. If we extrapolate
our results to $\Gamma=7$, we obtain the  value
$\Delta S = 0.35 k_B/{\rm ~ per ~ vortex ~ per ~ layer}$, in
very good agreement with experimental data on YBCO. This estimate is again in
surprisingly good agreement with, and only slightly larger
than, the result obtained by Hetzel {\it el.} for the uniformly frustrated
$3D XY$-model \cite{Hetzel:L92} with a considerably more sophisticated
technique, but where the effect of anisotropy was not fully accounted for. \\
There is also a weak specific heat anomaly at $T=T_z$, associated with the
loss of phase-coherence in the BCS order parameter along the field direction.
What this specific heat anomaly conceivably could show, is that there is a
phase-transition inside the flux-line liquid phase, from a low-temperature
flux-line liquid phase, to a high-temperature flux-line liquid phase, as
suggested by Feigel'man {\it et al.} \cite{Feigelman:B93}. In
Ref.  \cite{Feigelman:B93},  the low-temperature phase is suggested to
correspond to a flux-line liquid with no entanglement
of flux-lines, while the high-temperature flux-line liquid phase is suggested
to correspond to  a flux-line liquid with entanglement. In the language of
the $2D$ boson-analogy \cite{DRN:L88}, the former would correspond to a
normal Bose-liquid, while the latter would correspond to a superfluid
Bose-liquid, the two being separated by a genuine phase-transition. \\
{\it Note however, that the temperature $T_z$ goes down with increasing
system size, apparently with no sign of saturation, approaching $T_m$ from
above}. What this strongly indicates, is that the portion of the
phase-diagram with a flux-line liquid phase with an apparent intact
phase-coherence in the BCS order parameter across the sample parallel to
the magnetic field, will vanish in the thermodynamic limit. \\
The question remains as to what the character of the flux-line liquid
phase with no phase coherence along the magnetic field, is. We have
addressed the issue of whether the transition at $T_z \to T_m$ results in
a flux-line liquid with well-defined flux lines, by considering the amount
of flux-line cutting and intersectioning of flux-lines with closed vortex
loops, that takes place below and above the  temperature $T_z \to T_m$.

\subsection{Flux-line cutting}
\begin{figure}[htbp]
\psfig{figure=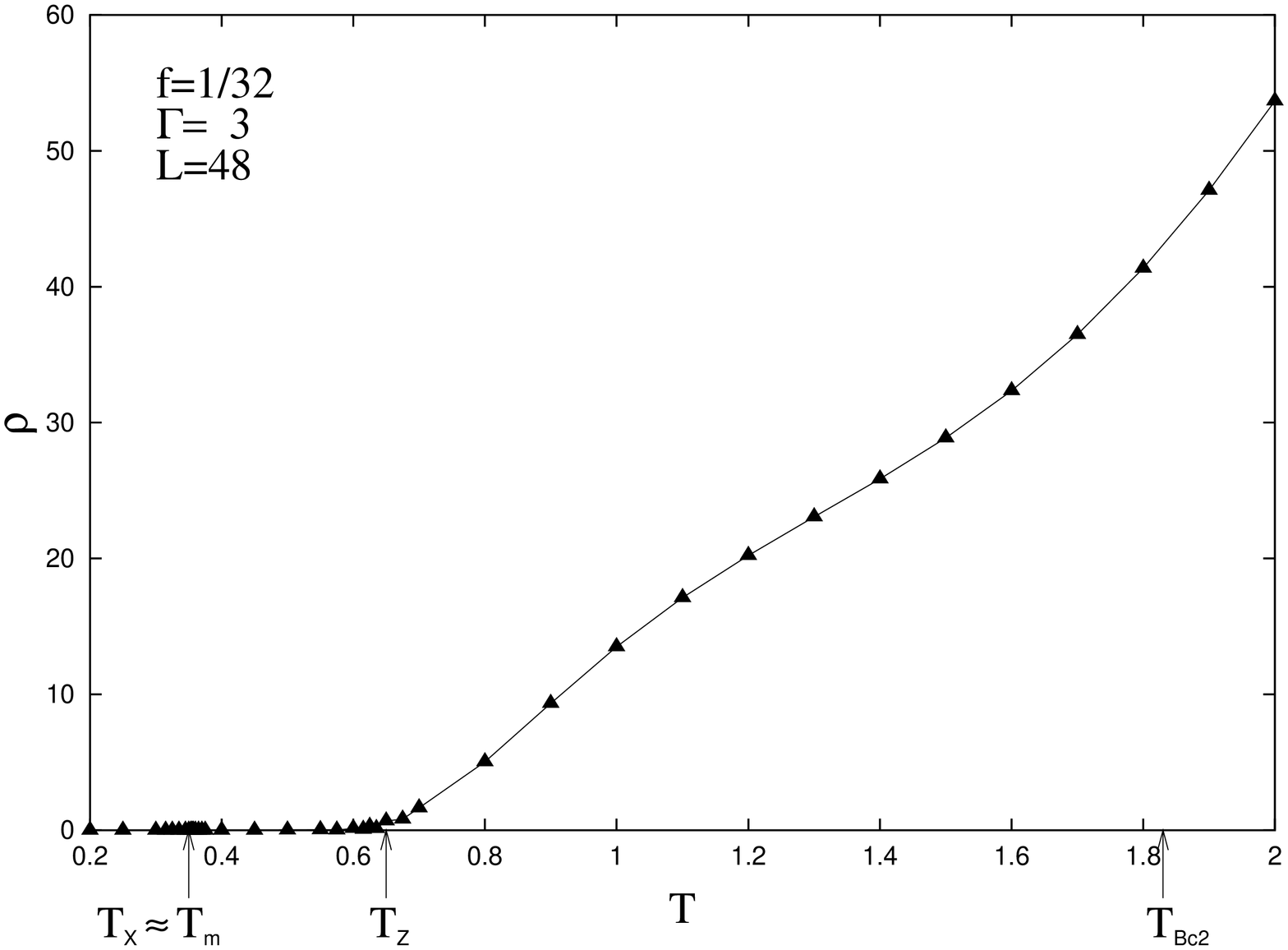,height=8cm,width=8cm,angle=270}
{\small FIG.~\ref{B32A3.Rho}. The number of  vortex-intersection events
	per flux-line $\rho$, for $f=1/32$, $\Gamma=3$, and $L=48$.}
\refstepcounter{figure}
\label{B32A3.Rho}
\end{figure}

A  flux-line liquid with large amounts of flux-line cutting events, is not
likely to be able to sustain a heavily entangled vortex configuration with
well defined flux lines. The amount of flux-line intersectioning, $\rho$
Eq. (\ref{rho}) is shown in Fig.  \ref{B32A3.Rho}.
As we see, $\rho$ increases sharply from zero at $T=T_z$, and continues
to increase monotonically as a function of temperature. Flux-line cutting
is an efficient way of disentangling flux lines, and the large values of
$\rho$ suggest that in the flux-line liquid phase, above $T_z$, the
flux-line liquid is incapable of sustaining an entangled
configuration. Hence, the loss of phase-coherence along the direction of
the magnetic field essentially is due to intersectioning between flux-lines,
and between flux lines and vortex-loops, with associated massive flux-line
recombinations. Recall that $T=T_z$ is also the temperature at which the
number of closed vortex-loops of diameter equal to the magnetic length in
the problem starts scaling with $L_z$, and it is natural to associate the
increase in $\rho$ with this limited proliferation of vortex-loops in
a finite field. Under such circumstances, a world-line picture of $2D$
bosons appears unlikely to be a particularly useful analogy to the flux-line
liquid system.

\section{Summary}
As discussed in section II.C, there can only be one single $3D$
phase-transition in the superconductor in zero field, regardless of the
anisotropy as long as this is finite, $\Gamma < \infty$. \\
We have shown that in zero field, the superconductor-normal metal
transition is due to a vortex-loop transition analogous to that first
suggested to occur in neutral superfluids, such as $He^4$, by Onsager
\cite{Onsager:Feynman}.  The transition is characterized by a vanishing
vortex-loop line tension, giving a loop-distribution function which decays as
$r^{-\alpha}$ at the transition, where $r$ is the radius of the loop. Below
$T_c$, the decay of the loop-distribution function clearly
appears to be faster-than-algebraic. If the
anisotropy is increased the vortex-loop unbinding temperature is reduced, but
the transition remains $3D$. Close to the transition the system is
isotropized due to an upward renormalization  of the interplane coupling as
a consequence of a diverging superconducting coherence length at the
transition.  \\
An important point is that in finite magnetic fields, the situation is
qualitatively different. In this case, the magnetic length of the vortex
system, i.e. the average distance between the flux-lines, cuts off the
upward renormalization of the inter-plane coupling
\cite{Tesanovic:B95,Nguyen:L96}. Note that the magnetic length only sets
a new length scale for the phase-coherence {\it as long as the flux-lines
actually fluctuate}. In the ground state, with no fluctuations in the
Abrikosov vortex lattice, the zero-field arguments apply \cite{Nguyen:L96}. \\
{\it Therefore, the perfect Abrikosov vortex lattice has no effect on the
vortex-loop blowout transition, since the phase-coherence length is
essentially infinite for this case. Consequently, for the case
$\lambda = \infty$, one cannot have a vortex-loop blowout transition
{\it far below} the flux-line lattice melting temperature.} \\
As soon as the flux lines start to fluctuate appreciably, which only happens
very close to the melting transition due to the first order character of
the transition, the resulting much smaller cutoff on the renormalization of
the inter-plane coupling facilitates a vortex loop blowout virtually at
the same temperature as the flux lines start to fluctuate. We emphasize that
these statements apply to the case of total suppression of gauge-fluctuations,
or $\lambda=\infty$. \\
Even if vortex-loops do not exist on all length scales in the problem
at the flux-line lattice melting temperature, they will seriously affect the
flux-line liquid phase provided that they exist on length scales up to the
magnetic length of the problem. {\it From our simulations, we have found
that this is always the case in the thermodynamic limit.}: The number
of such closed non-field induced closed vortex-loops scales with the
thickness of the sample, $L_z$, whereas the number of field-induced
flux-lines obviously does not. Therefore there is an infinite number of
closed vortex-loops with a diameter equal to the magnetic length, per
flux-line. Hence, in the liquid phase flux lines cannot be considered as
well defined entities. They are well-defined entities in the
{\it lattice} case. \\
When a flux-line lattice melts, the molten phase is an {\it incoherent
vortex-liquid} characterized by loss of phase-coherence along the
direction of the magnetic field. This loss of phase-coherence is associated
with a proliferation of closed vortex loops and massive flux-line cutting
and recombination. Such a vortex-liquid phase is unlikely to sustain
heavily entangled vortex-configurations. \\
{\it Note added}: After this work was completed, we received works from
Hu, Miyashita, and Tachiki \cite{Tachiki:97}, as well as from Koshelev
\cite{Koshelev:97}, that support the conclusion that $T_z=T_m$
when $\lambda = \infty$, at least in the field regime considered, i.e $f=1/25$
\cite{Tachiki:97} and $f=1/36$ \cite{Koshelev:97}.

\section{Acknowledgments}
Support from the Research Council of Norway (Norges Forskningsr{\aa}d)
under Grants No. 110566/410, No. 110569/410, as well as a grant for
computing time under the Program for Super-computing, is gratefully
acknowledged. We thank G. Blatter, C. Dasgupta, {\O}. Fischer,
V. B. Geshkenbein, R. E. Hetzel, A. Junod, A. E. Koshelev,
H. Nordborg, M. Tachiki, S. Teitel, Z. Te{\v s}anovi{\'c}, N. C. Yeh,
and C. Yu for discussions. J. Amundsen is acknowledged for assistance in
optimizing our computer codes for use on the Cray T3E.

\appendix
\section{Helicity modulus in terms of phase-variables}
In this appendix we derive the expression for the helicity modulus
$\Upsilon_\mu$ (Eq.~\ref{helmod}), for the uniformly frustrated
{\em anisotropic} Villain model. \\
The effective Hamiltonian for the uniformly frustrated anisotropic
Villain model is
\begin{eqnarray}
   H_v \{\theta'(\vec{r})\} = J_0 \sum_{\vec{r},\nu= x,y,z}
      V_\nu[\theta'(\vec{r}+\hat{e_\nu}) - \theta'(\vec{r}) - A_\nu(\vec{r})]
\nonumber \\
   V_\nu(\chi) = -\frac{k_B T}{J_0}
      \ln \left \{ \sum_{m = -\infty}^{\infty}
      \exp \left [-\frac{J_0 \alpha_\nu}{2k_B T}
      (\chi - 2 \pi m)^2 \right ] \right \}.
\label{A1VillainPot}
\end{eqnarray}
Here, $J_0$, $A_\nu(\vec{r})$ and $\alpha_\nu$ are defined
in the text, $\hat{e}_\nu$ is the unit vector for the $\hat{\nu}$-axis.
In Eq.~\ref{A1VillainPot}, $\theta'(\vec{r})$ is the phase of the
complex superconducting order parameter. $\{\theta'(\vec{r})\}$ denotes
the functional of $\theta'(\vec{r})$.
We now apply boundary conditions such that the phase across the system
in the $\hat{\mu}$-direction is twisted by an amount $L \delta$. If
$\hat{\mu}=\hat{z}$, for example, we have the following phase twist
\begin{eqnarray}
   \theta'(x,y,z=L) - \theta'(x,y,z=0) = L \delta .
\label{Bound.Theta}
\end{eqnarray}
Now we define a set of new phase variables
\begin{eqnarray*}
   \theta(\vec{r}) = \theta'(\vec{r}) - (\vec{r} \cdot \hat{e}_\mu) \delta .
\end{eqnarray*}
If the phase variables $\theta'(\vec{r})$ obey the twisted boundary conditions
in Eq.~\ref{Bound.Theta}, the new phase variables $\theta(\vec{r})$ obey
the following periodic boundary conditions
\begin{eqnarray}
   \theta(x,y,z=L) - \theta(x,y,z=0) = 0.
\label{Bound.ThetaP}
\end{eqnarray}
The Hamiltonian for the uniformly frustrated anisotropic Villain model
in terms of the new phase variables is
\begin{eqnarray*}
   H_v\{\theta,\delta\}=J_0 \! \! \! \! \sum_{\vec{r},\nu=x,y,z}
      \! \! \! \!
      V_\nu[\theta(\vec{r}+\hat{e_\nu}) - \theta(\vec{r}) - A_\nu(\vec{r})
      + (\hat{e}_\nu \cdot \hat{e}_\mu) \delta].
\end{eqnarray*}
The partition function in terms of the new phase variables is
\begin{eqnarray*}
   Z(\delta) = \sum_{\{\theta(\vec{r})\}}
      e^{-\frac{H_v\{\theta(\vec{r}),\delta\}}{k_B T}},
\end{eqnarray*}
where we only sum over configurations $\{\theta(\vec{r})\}$ that satisfy
periodic boundary conditions. The total free energy is
\begin{eqnarray*}
   F(\delta) = - k_B T~\ln Z(\delta).
\end{eqnarray*}
The helicity modulus $\Upsilon_\mu$ is defined as the second derivative of
the free energy with respect to a phase twist across the sample in the
$\hat{\mu}$-direction. Thus,
\begin{eqnarray*}
   \Upsilon_\mu \equiv \frac{1}{L^3} \left( \frac{\partial^2 F(\delta)}
        {\partial \delta^2} \right)_{\delta = 0}.
\end{eqnarray*}
Using the definition
\begin{eqnarray*}
   \chi & \equiv & \theta(\vec{r}+\hat{e}_\nu) - \theta(\vec{r})
   - A_\nu(\vec{r}) + (\hat{e}_\nu \cdot \hat{e}_\mu) \delta , \\
   \frac{\partial \chi}{\partial \delta} & = & \hat{e}_\nu \cdot \hat{e}_\mu,
\end{eqnarray*}
we can write $\Upsilon_\mu$ in the following form
\begin{eqnarray*}
   \Upsilon_\mu =  \frac{J_0^2}{L^3} \frac{1}{k_B T} & \left (
      \frac{{\displaystyle \sum}_{ \{ \theta(\vec{r})\}}
      \left[\sum_{\vec{r},\nu} \frac{\partial}{\partial \chi}
      V_\nu(\chi) (\hat{e}_\nu \cdot \hat{e}_\mu) \right]
      e^{-\frac{H_v\{\theta(\vec{r})\}}{k_B T}}}
      {Z(\delta=0)}\right )^2
\\
     +\frac{J_0}{L^3}~~~~~~ &
      \frac{{\displaystyle \sum}_{ \{ \theta(\vec{r})\}}
      \left[\sum_{\vec{r},\nu} \frac{\partial^2}{\partial \chi^2}
      V_\nu(\chi) (\hat{e}_\nu \cdot \hat{e}_\mu)^2 \right]
      e^{-\frac{H_v\{\theta(\vec{r})\}}{k_B T}}}
      {Z(\delta=0)}
\\
      -  \frac{J_0^2}{L^3} \frac{1}{k_B T} &
      \frac{{\displaystyle \sum}_{ \{ \theta(\vec{r})\}}
      \left[\sum_{\vec{r},\nu} \frac{\partial}{\partial \chi}
      V_\nu(\chi) (\hat{e}_\nu \cdot \hat{e}_\mu) \right]^2
      e^{-\frac{H_v\{\theta(\vec{r})\}}{k_B T}}}
      {Z(\delta=0)},
\end{eqnarray*}
which is Eq.~\ref{helmod} written in a more explicit form.
The summations over the configurations $\{\theta({\vec{r}_i})\}$ are
restricted to configurations satisfying periodic boundary conditions.

\section{Helicity modulus in terms of vortex segments correlations}
In this appendix we derive the helicity modulus for the
anisotropic LSM $\Upsilon_\mu(\vec{k})$, expressed in terms of vortex
density-density correlation functions \cite{Nguyen:L96}, $\mu = (x,y,z)$.\\
In the continuum limit, the effective Hamiltonian for the anisotropic
LSM can be written as
\begin{eqnarray}
	H\{\vec{v},\vec{f}\} = \frac{J_1}{2}
	   \! \int \! d^3r \! \left[ \vec{v}(\vec{r}) \cdot
	   \stackrel{\leftrightarrow}{M} \cdot \vec{v}(\vec{r}) +
	   2 \pi \lambda^2 \vec{f}(\vec{r}) \cdot \vec{f}(\vec{r}) \right ],
\label{H_LSM}
\end{eqnarray}
the energy per unit length $J_1=\Phi_0^2/16\pi^{3}\lambda_a^2$.
Here, $\vec{v}(\vec{r})$ is the super-fluid velocity,
\begin{eqnarray*}
	\vec{v}(\vec{r}) = \vec{\nabla} \theta (\vec{r}) - \vec{A}(\vec{r}),
\end{eqnarray*}
$\theta(\vec{r})$ is the phase of the complex superconducting order parameter,
$\Phi_0 \vec{A}(\vec{r})/2 \pi$ is the vector potential, $\Phi_0$ is the
flux quantum. In Eq.~\ref{H_LSM}, $\stackrel{\leftrightarrow}{M}$ is the
anisotropic mass tensor,
\begin{eqnarray*}
	\stackrel{\leftrightarrow}{M} = \left  [
	   \begin{array}{ccc}
		~1~ & ~0~ & ~0~ \\
		~0~ & ~1~ & ~0~ \\
		~0~ & ~0~ & ~\Gamma^2~
	   \end{array}                  \right ],
\end{eqnarray*}
describing uniaxial $\hat c$-anisotropy. The anisotropy parameter
$\Gamma = \lambda_c/\lambda_a$, $\lambda_a$ and $\lambda_c$ are explained
in the text. In Eq.~\ref{H_LSM}, $\vec{f}(\vec{r})$ is the local density
of the magnetic flux quanta,
\begin{eqnarray*}
	\vec{f}(\vec{r}) = \frac {\vec{B}(\vec{r})} {\Phi_0} =
	   \frac {1} {2 \pi} \vec{\nabla} \times \vec{A}(\vec{r}).
\end{eqnarray*}
Here, $\vec{B}(\vec{r})$ is the local magnetic induction.
In evaluating Eq.~\ref{H_LSM}, the integration must be cut off at the core
of the vortex segments, so that the energy stays finite. \\
The partition function $Z$ is computed averaging over independently
fluctuating $\vec{v}(\vec{r})$ and $\vec{f}(\vec{r})$, subject to the
constraint $\langle \vec{f}(\vec{r}) \rangle = f \hat{e}_z$ for a constant
uniform average magnetic induction $B \hat{e}_z$, where
$\langle ... \rangle$ denotes a thermal average, and $\hat{e}_\mu$ is
the unit vector along the $\hat{\mu}$-axis. \\
Using the Fourier transform
\begin{eqnarray*}
	\tilde{v}(\vec{k}) = \int d^3r~ e^{i \vec{k} \cdot \vec{r}}
			                \vec{v}(\vec{r}),
\end{eqnarray*}
the Hamiltonian in Eq.~\ref{H_LSM} can be written as
\begin{eqnarray}
	H\{\tilde{v}(\vec{k}),\tilde{f}(\vec{k})\} =
	    \frac{J_1}{2 V} \sum_{\vec{k}} H'(\vec{k}), \nonumber \\
	H'(\vec{k}) =
	   \left [ \tilde{v}(\vec{k}) \cdot \stackrel{\leftrightarrow}{M}
	   \cdot \tilde{v}(-\vec{k}) + 2\pi \lambda^2 \tilde{f}
	   (\vec{k}) \cdot \tilde{f}(-\vec{k}) \right ],
\label{H.k}
\end{eqnarray}
where V is the volume of the system.
An applied twist in the phase of the superconducting order parameter,
along the $\hat{\mu}$-direction, as considered in the previous Appendix,
corresponds to a change in the super-fluid velocity
\begin{eqnarray*}
	\tilde{v}(\vec{k}) ~\rightarrow~ \tilde{v}(\vec{k}) +
	\delta v (\vec{k}) \hat{e}_\mu.
\end{eqnarray*}
In terms of velocities, therefore, the helicity modulus
$\Upsilon_\mu(\vec{k})$ is defined as the second
derivative of the free energy \mbox{$F = -k_B T \ln Z$} with respect to a
change in the superfluid velocity  along the $\hat{\mu}$-direction. Hence,

\bleq
\begin{eqnarray}
\nonumber
	\Upsilon_\mu(\vec{k}) & \equiv & \left. \frac {\partial^2 F}
	   {\partial[\delta v(\vec{k})] \partial [\delta v (- \vec{k})]}
	   \right |_{\delta v(\vec{k})=\delta v (-\vec{k}) = 0}
\\
	   & = & \frac{J_1}{V}
	   \left ( \hat{e}_\mu \cdot \stackrel{\leftrightarrow}{M}
	   \cdot \hat{e}_\mu - \frac{J_1}{k_B T V} \left \langle
	   [\tilde{v}( \vec{k}) \cdot \stackrel{\leftrightarrow}{M}
	   \cdot \hat{e}_\mu] [\hat{e}_\mu \cdot \stackrel{\leftrightarrow}{M}
	   \cdot \tilde{v}(-\vec{k})] \right \rangle_0 \right ).
\label{Upsilon_v}
\end{eqnarray}
\eleq

The subscript 0 denotes the unperturbed system with no applied phase twist,
$\delta v (\vec{k}) = 0$. In Eq.~\ref{Upsilon_v}, we have used
$\langle \tilde{v}(\vec{k}) \rangle_0 = \langle \tilde{v}(-\vec{k})
\rangle_0 = 0$, since the Hamiltonian Eq.~\ref{H.k} contains only
quadratic terms like $[v_\nu(\vec{k}) v_\nu(-\vec{k})]$  $(\nu=x,y,z)$.
Because of the symmetry in which $\vec{\nabla} \theta$ and $\vec{A}$ enter
$\vec{v}(\vec{r})$, $\Upsilon_\mu(\vec{k})$ can also
be interpreted as the linear response coefficient of the supercurrent
induced by a perturbation in the vector potential \cite{Teitel:L94},
\begin{eqnarray*}
	j_\mu(\vec{k}) = - \Upsilon_\mu(\vec{k}) \delta A_\mu(\vec{k}).
\end{eqnarray*}
There are three interesting helicity moduli to be considered:
(1) $\Upsilon_x(k \hat{e}_y)$, which is the energy cost corresponding to a
compressional perturbation of the flux-line system.
(2) $\Upsilon_x(k \hat{e}_z)$, which is the energy cost corresponding to a
tilting perturbation of the flux-line system.
(3) $\Upsilon_z(k \hat{e}_x)$, which is the energy cost corresponding to a
shearing of the flux-line system. \\
To find an expression for $\Upsilon_\mu(k\hat{e}_\nu)$ $(\mu \not = \nu)$
expressed in terms of vortex segments density correlations,
we need to write the parts of H containing $k\hat{e}_\nu$ in a diagonal form.
Defining the vortex segment density
\begin{eqnarray*}
	\vec{n}(\vec{r}) = \frac{1}{2 \pi} \vec{\nabla} \times
	   \vec{\nabla} \theta,
\end{eqnarray*}
we can write the super-fluid velocity in the following gauge-invariant form
\cite{Teitel:L94}
\begin{eqnarray}
	\tilde{v}(\vec{k}) = 2 \pi i \left ( \vec{k} \chi(\vec{k})
	   + \frac{\vec{k} \times [\vec{n}(\vec{k})-\vec{f}(\vec{k})]}
	          {k^2} \right ).
\label{v.k}
\end{eqnarray}
Here, $\chi(\vec{r})$ is a smooth scalar function which describes the
longitudinal part of $\tilde{v}(\vec{k})$. The transverse part of
$\tilde{v}(\vec{k})$ is determined by $\vec{\nabla} \times \vec{v}(\vec{r})
= 2 \pi [\vec{n}(\vec{r}) - \vec{f}(\vec{r})]$. \\
Substituting Eq.~\ref{v.k} into $H'(\vec{k})$ (Eq.\ref{H.k}), we get
the following diagonal form for $H'(\vec{k})$ for the case of
$\vec{k} = k \hat{e}_\nu$

\bleq
\begin{eqnarray}
	\frac{H'(k \hat{e}_x)}{4 \pi^2} = ~~~~~
	   k^2 \chi (k \hat{e}_x) \chi (-k \hat{e}_x)          &+&
	   A_1(k) n_y(k \hat{e}_x) n_y(-k \hat{e}_x)            +
	   A_2(k) n_z(k \hat{e}_x) n_z(-k \hat{e}_x)            +
\nonumber \\ & &
	   B_1(k) \delta f_y(k \hat{e}_x) \delta f_y(-k \hat{e}_x) +
	   B_2(k) \delta f_z(k \hat{e}_x) \delta f_z(-k \hat{e}_x),
\nonumber \\
	\frac{H'(k \hat{e}_y)}{4 \pi^2} = ~~~~~
	   k^2 \chi (k \hat{e}_y) \chi (-k \hat{e}_y)          &+&
	   A_1(k) n_x(k \hat{e}_y) n_x(-k \hat{e}_y)            +
	   A_2(k) n_z(k \hat{e}_y) n_z(-k \hat{e}_y)            +
\nonumber \\ & &
	   B_1(k) \delta f_x(k \hat{e}_y) \delta f_x(-k \hat{e}_y) +
	   B_2(k) \delta f_z(k \hat{e}_y) \delta f_z(-k \hat{e}_y),
\nonumber \\
	\frac{H'(k \hat{e}_z)}{4 \pi^2} =
	   (k/\Gamma)^2 \chi (k \hat{e}_z) \chi (-k \hat{e}_z) &+&
	   A_2(k) n_x(k \hat{e}_z) n_x(-k \hat{e}_z)            +
	   A_2(k) n_y(k \hat{e}_z) n_y(-k \hat{e}_z)            +
\nonumber \\ & &
	   B_2(k) \delta f_x(k \hat{e}_z) \delta f_x(-k \hat{e}_z) +
	   B_2(k) \delta f_y(k \hat{e}_z) \delta f_y(-k \hat{e}_z).
\label{H.k.diagonal}
\end{eqnarray}
Here,
\begin{eqnarray*}
	A_1(k) = \frac{\lambda^2}{1 + \Gamma^2 \lambda^2 q^2},~~~~~~~~
	A_2(k) = \frac{\lambda^2}{1 +          \lambda^2 q^2}, \\
	B_1(k) = \frac{1 + \Gamma^2 \lambda^2 q^2}{\Gamma^2 q^2},~~~~~~~~
	B_2(k) = \frac{1 +          \lambda^2 q^2}{        q^2}.
\end{eqnarray*}
In Eq.~\ref{H.k.diagonal},
$\delta \tilde{f}(\vec{k}) = \tilde{f}(\vec{k}) - \tilde{f}^0(\vec{k})$
is the fluctuation of the magnetic flux density
away from the value $\tilde{f}^0(\vec{k})$ minimizing the Hamiltonian
for a given vortex segments configuration \cite{Brandt:JLT77}.
\begin{eqnarray*}
	\tilde{f}^0(\vec{k}) = \frac{\tilde{n}(\vec{k})}{1+\lambda^2 k^2} -
	   \frac{(\Gamma^2 - 1)[\tilde{n}(\vec{k}) \cdot \vec{q}]
	         ~ \lambda^2 \vec{q}}
                {1+\lambda^2 k^2 + (\Gamma^2-1) \lambda^2 q^2},
\end{eqnarray*}
where $\vec{q}=\vec{k} \times \hat{z}$.
To compute the partition function Z, we should sum over: (1) all
smooth functions $\chi(\vec{r})$, (2) all $\tilde{n}(\vec{k})$ that satisfy
$\vec{k} \cdot \tilde{n}(\vec{k}) = 0$, and (3) all
$\delta \tilde{f}(\vec{k})$ that
satisfy $\vec{k} \cdot \delta \tilde{f}(\vec{k}) = 0$. The constraints
(2) and (3) come from the restriction of no divergence in the vortex segments
density and no divergence in the local magnetic field induction.
Substituting Eq.~\ref{v.k} into Eq.~\ref{Upsilon_v}, and using the the
Hamiltonian Eq.~\ref{H.k.diagonal} to evaluate the average over $\chi(\vec{k})$
and $\delta \tilde{f}(\vec{k})$. For the case of $\vec{k} = k \hat{e}_\nu$
we obtain
\begin{eqnarray}
	\Upsilon_\mu(k \hat{e}_\nu) =
	   \frac{J_1}{V}
	   \frac{\lambda^2 k^2}
	        {1+[1+\delta_{\mu,z}(\Gamma^2-1)]\lambda^2 k^2}
	   \left ( 1 - \frac{4 \pi J_1}{k_B T V}
	       \frac{\lambda^2 ~ \langle n_\sigma(k \hat{e}_\nu)
                     n_\sigma(-k \hat{e}_\nu) \rangle_0}
	            {1+[1+\delta_{\mu,z}(\Gamma^2-1)]\lambda^2 k^2}
           \right ),
\label{Helicity.Modulus.Continum}
\end{eqnarray}
$(\mu, \nu, \sigma)$ are cyclic permutation of (x, y, z). The generalization
of Eq.~\ref{Helicity.Modulus.Continum} to a lattice superconductor is
\begin{eqnarray}
	\Upsilon_\mu(k \hat{e}_\nu) =
	   \frac{J_0}{L^3}
	   \frac{(\lambda/d)^2 Q^2}{1+[1+\delta_{\mu,z}
	        (\Gamma^2-1)](\lambda/d)^2 Q^2}
	   \left ( 1 - \frac{4 \pi J_0}{k_B T L^3}
	       \frac{(\lambda/d)^2 ~ \langle n_\sigma(k \hat{e}_\nu)
		     n_\sigma(-k \hat{e}_\nu) \rangle_0}
	            {1+[1+\delta_{\mu,z}(\Gamma^2-1)](\lambda/d)^2 Q^2}
           \right ),
\label{Helicity.Modulus.Lattice}
\end{eqnarray}
where $Q_\mu = 2 sin(k_\mu d/2)$, $Q^2 = \sum_\mu Q_\mu^2$, $k_\mu$ is the
$\mu$-component of $\vec{k}$, and d is the lattice constant of the
(simple cubic) underlying numerical lattice.
\eleq

\end{multicols}
\end{document}